\documentclass[
groupedaddress,
nofootinbib,
amsmath,
amssymb,
twocolumn,
prx,
floatfix,
]{revtex4-2}
\usepackage[T1]{fontenc}

\usepackage{amsfonts}

\usepackage{graphicx}
\usepackage{dcolumn}
\usepackage{bm}

\usepackage{svg}
\usepackage{algorithm}
\usepackage{algpseudocode}
\usepackage{nth}

\usepackage{tikz}
\usetikzlibrary{quantikz2}

\makeatletter
\newcommand{\superimpose}[3][\mathord]{#1{\mathpalette\superimpose@{{#2}{#3}}}}
\newcommand{\superimpose@}[2]{\superimpose@@{#1}#2}
\newcommand{\superimpose@@}[3]{%
  \ooalign{%
    \hfil$\m@th#1#2$\hfil\cr
    \hfil$\m@th#1#3$\hfil\cr
  }%
}
\makeatother

\newcommand{\ketbra}[2]{\left\vert#1\right\rangle\left\langle#2\right\vert}
\newcommand{\pis}{{+^{\small \mathcal V}}}

\begin{document}

\preprint{APS/123-QED}

\title{Quantum Subspace Correction for Constraints}

\author{Kelly Ann Pawlak}
\email{kellyp@atom-computing.com}
\affiliation{Atom Computing, Berkeley, CA}

\author{Jeffrey M. Epstein}
\affiliation{Atom Computing, Berkeley, CA}

\author{Daniel Crow}
\affiliation{Atom Computing, Berkeley, CA}

\author{Srilekha Gandhari}
\affiliation{Atom Computing, Berkeley, CA}

\author{Ming Li}
\affiliation{Atom Computing, Berkeley, CA}

\author{Thomas C. Bohdanowicz}
\affiliation{Atom Computing, Berkeley, CA}

\author{Jonathan King}
\email{jonathanking@atom-computing.com}
\affiliation{Atom Computing, Berkeley, CA}

\date{\today}

\begin{abstract}
We demonstrate that it is possible to construct operators that stabilize the constraint-satisfying subspaces of computational problems in their Ising representations. We provide an explicit recipe to construct unitaries and associated measurements given a set of constraints. The stabilizer measurements allow the detection of constraint violations, and provide a route to recovery back into the constrained subspace. We call this technique ``quantum subspace correction". As an example, we explicitly investigate the stabilizers using the simplest local constraint subspace: Independent Set. We find an algorithm that is guaranteed to produce a perfect uniform or weighted distribution over all constraint-satisfying states when paired with a stopping condition: a quantum analogue of partial rejection sampling. The stopping condition can be modified for sub-graph approximations. We show that it can prepare exact Gibbs distributions on $d-$regular graphs below a critical hardness $\lambda_d^*$ in sub-linear time. Finally, we look at a potential use of quantum subspace correction for fault-tolerant depth-reduction. In particular we investigate how the technique detects and recovers errors induced by Trotterization in preparing maximum independent set using an adiabatic state preparation algorithm. 
\end{abstract}

\maketitle
\newpage

\section{Introduction}
Neutral atoms are an emerging platform for quantum computation. This architecture can feature exceptionally long coherence times on the order of 40 seconds\cite{barnes2022assembly}, high connectivity, and mid-circuit qubit rearrangement, native multi-qubit gates,\cite{levine2019parallel, bluvstein2022quantum, graham2022multi, ma2023highfidelity} and has recently yielded demonstrations of mid-circuit measurement\cite{norcia2023mid, graham2023mid, lis2023mid, PhysRevLett.129.203602}, ultimately promising near-term feed-forward and error-correction capabilities.

Given the abilities of these new systems, we are motivated to study algorithms suited to their strengths in both the fault-tolerant (FT) and pre-FT regimes. To date, development of pre-FT algorithms has largely focused on variational algorithms, which aim to maximize the utility of noisy qubits by classically optimizing parameterized circuits\cite{bharti2022noisy}. Variational algorithms typically require extensive sampling, which presents a particular challenge for neutral atom QPUs with relatively long readout times\cite{norcia2023mid}. Recent work attempts to address this challenge in terms of reducing readout times\cite{PhysRevLett.129.203602, dhordjevic2021entanglement} or optimization with limited sampling\cite{polloreno2023qaoa}, but here we focus on an alternative approach that takes particular advantage of the already-demonstrated long coherence times and mid-circuit measurement available in neutral atom quantum computers.

\begin{figure}
    \centering
    \includegraphics[width=2.4in]{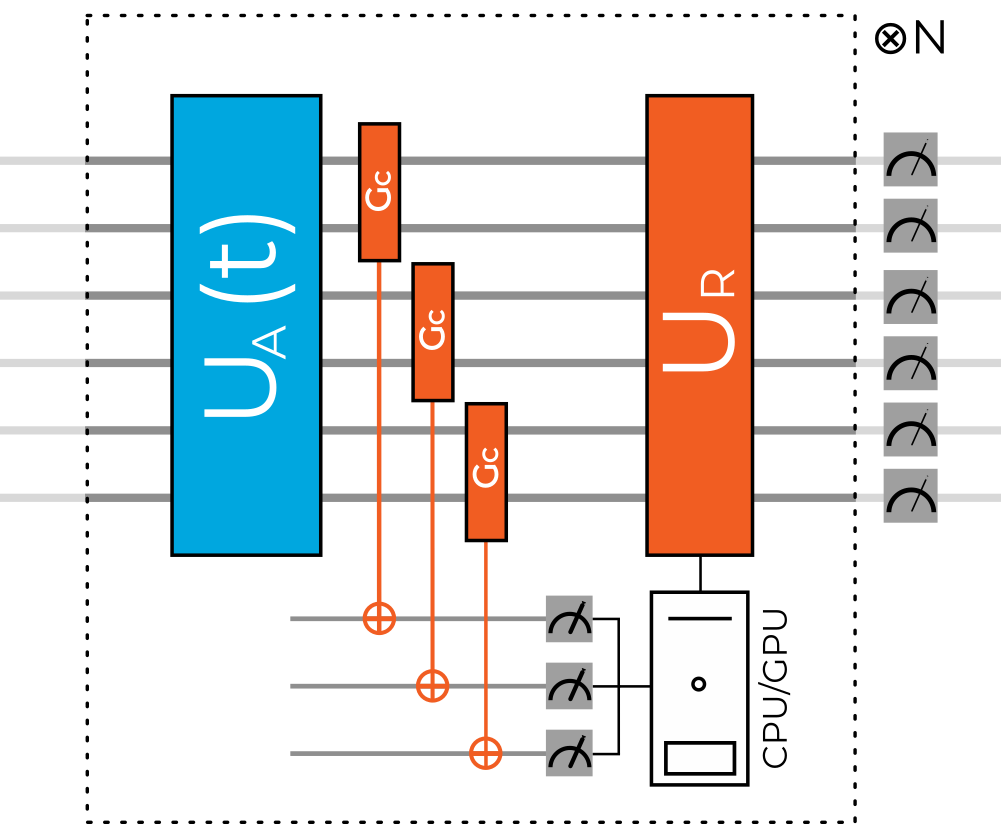}
    \caption{On atomic quantum computing hardware, coherence time (QEC cycle times), qubit count, rearrangement, and native multiqubit gates can deeply expand the non-trivial classical computations possible within the coherent execution of a circuit. These abilities have been, thus far, underutilized in many popular algorithm development approaches in the past decade where such capabilities were not practical.}
    \label{fig:NAQC capabilities}
\end{figure}

More recently, a small wave of algorithm development has targeted ion-trap computers, which can feature design elements such as all-to-all connectivity or feed-forward (also called adaptive) circuit capability. In particular, the work presented in this paper has been inspired by the use of adaptive circuits to prepare ground states of long-range entangled states in shorter depth than possible with unitaries alone \cite{foss2023experimental, bravyi2022adaptive}. Especially in the case of the toric code ground state of \cite{foss2023experimental}, the perspective can be understood as retooling of error correction techniques to solve computational, rather than quantum control, problems. 

In this publication, we detail a new class of hybrid algorithms intended for neutral atom quantum computers (NAQC)  centered on a technique we call ``quantum subspace correction" (QSC) that explicitly checks the subspace of a quantum register via strong projective measurements and performs a recovery on violating portions of the state.

QSC is built from the ideas of Quantum Error Correction (QEC). For a given subspace, \textit{e.g.} a constrained subspace, one constructs a set of generalized stabilizers whose syndromes determine if the qubit wavefunction resides in the desired subspace or not. The information obtained from the readout of all stabilizer measurements can be aggregated and collectively used to prepare a recovery back into the intended subspace. We believe that there are a number of promising uses of this technique. Most importantly, from a practical standpoint, QSC specifically utilizes the theoretical framework and practical workflow of full error correction: in this way, the search for early use-cases and new algorithm development becomes intrinsically aligned with long term FTQC engineering goals in NAQCs, rather than a detour. Moreover this technique, as we show, leads to new and potentially interesting early-FT and possibly enduring FT algorithms.

As a first application of QSC, we specifically consider the topic of constraint satisfaction, either in effort to construct satisfying solutions to a set of constraints or as part of a larger optimization problem or dynamical simulation. In  Section \ref{sec:IS-construction}, we focus on the problem of Independent Set (IS), in  having the simplest constraint structure, providing explicit stabilizers for this problem in the main text. Some additional examples of stabilizers for other common constraints are to be provided in \ref{app:detections}. We outline two use-cases. In Section \ref{sec: Exact Dist}, we explore exact distribution preparation using QSC as a tool to perform Quantum Partial Rejection Sampling. We investigate the features of this algorithm in depth-bounded and pre-fault tolerant use-cases. In Section \ref{sec: adiabatic}  we briefly look at some potential applications to constrained optimization by using QSC in tandem with a clever adiabatic algorithm for maximum independent set first developed by Ref. \cite{yu2021quantum}. Finally, in Section \ref{sec:discuss} we review the merits and limitations of some of these techniques, and chart out a path for future work.

\section{Quantum Subspace Correction for the Independent Set Constraint} \label{sec:IS-construction}
Due to its simplicity and ubiquity, in this section we will construct the tools for quantum subspace correction using the independent set (IS) constraint: 

\textbf{Definition (Independent Set):} \textit{Given an undirected graph \( \mathcal G = (V, E) \), an independent set comprises a subset of vertices \( I \subseteq V \) where no two vertices within \( I \) share an edge. Formally, for every \( u, v \in I \), \( (u, v) \notin E \).}

The widely-accepted Ising form of the independent set problem is detailed in \cite{lucas2014ising}, along with other common constraints that we consider in Appendix \ref{app:detections}. Here, each vertex in the graph corresponds to a qubit. When measured in the computational basis, the state of this qubit, which belongs to the set $\{0,1\}$, signifies either exclusion (0) or inclusion (1) of the vertex in the set. 

For any two qubits that represent vertices connected by an edge, a measurement of the $|11\rangle$ state indicates a violation of the independence constraint. In this way, we define an edge subspace that satisfies the independent set constraint, $\mathcal V_e$ as any superposition over the basis states $\{|00\rangle, |01\rangle, |10\rangle\}$. The orthogonal complement of $\mathcal V_e$ on each edge is $\bar{ \mathcal V}_e$ and supported by the $|11\rangle$ basis vector.

For each edge, one can construct a operator, $\hat S_e$ such that all $|\psi\rangle \in \mathcal V_e $ are in the $+1$ eigenspace of $\hat S_e$, and all $|\psi\rangle \in \bar{\mathcal V}_e $  are in the $-1$ eigenspace. This operator has the explicit form $\hat S_{(i,j)} =\frac{1}{2}(I + Z_i +Z_j -Z_iZ_j) $.

These operators have the following  properties:
\begin{enumerate}
    \item The operators $\{\hat S_e\}$ form a group: 
\begin{align*}
    &\text{if}\; \hat S_e |\psi\rangle = |\psi\rangle \;\mathrm{and}\; \hat S_{e'} |\psi\rangle = |\psi\rangle 
    \\
    &\text{then}\;    \hat S_{e}  \hat S_{e'} |\psi\rangle =|\psi\rangle
\end{align*}
\item The operators $\{\hat S_e\}$ are Abelian:
\begin{align*}
    &\text{if}\; \hat S_e |\psi\rangle = |\psi\rangle \;\mathrm{and}\; \hat S_{e'} |\psi\rangle = |\psi\rangle 
    \\
    &\text{then}\;    [\hat S_{e}, \hat S_{e'}] |\psi\rangle = 0 \; 
\end{align*}
\end{enumerate}

The joint $+1$ eigenspace under all $\hat S_e$ for $e \in E$ is:
\begin{equation}
    \prod_{e \in E}  \hat S_e |\psi\rangle = +1 |\psi\rangle\qquad \text{for } \; |\psi\rangle \in \mathcal V 
\end{equation}
Given the properties of $\{\hat S_{e}\}$, we can identify this set of operators as the stabilizer of $\mathcal V$ \cite{gottesman1997stabilizer, gottesman1996class}, which represents the global subspace of all valid independent sets on a graph $\mathcal G = (V, E)$. 

For a given graph $\mathcal G$ for which we are interested in states obeying IS, QSC should be applied to the relevant qubits. The result of measuring these stabilizers creates a violation graph which labels all edges where a violation of IS exists. Depending on the goal of the algorithm, a recovery can then be classically computed and implemented as unitary gates.

Measuring this stabilizer in practice is a straight-forward problem of finding the correct controlled unitary to implement onto an ancilla for a given set of constraints. Remarkably, the stabilizer for the independent set problem corresponds to \textit{a single Toffolli gate controlled on the edge qubits, acting on an ancilla}. A primitive of this gate (CCZ) is available as a native gate on neutral atom quantum computers due to the Rydberg interaction\cite{levine2019parallel, li2023one, henriet2020quantum, muller2009mesoscopic}. Furthermore, in a restricted Clifford$+T$ FTQC, Toffolli gates are a low level building block that require only $7$ $T-$gates and $8$ Clifford gates. 

For a fully parallel implementation, the number of ancillae should be equivalent to $|E|$, potentially making use of the large numbers of qubits available on NAQC hardware. However, it is possible to perform syndrome extraction with this stabilizing code using as few as one ancilla with repeated measurements and resets. In a FTQC, the stabilizing measurement rounds for QSC can be mixed in with the regular QEC cycle times, given that the error correction cycle is long enough to allow for additional classical computation.

In the next sections, we give some examples of what to do with these stabilizers. For a general objective, such as optimization, since the stabilizer only checks the constraint violation, and it is complicated to generate a list of all operations that could lead to any violation, it is typical that a recovery operation is not uniquely specified as in the case with QEC codes. Otherwise, due to the structure of the constraint space, a code on it is usually highly degenerate. For example, in the construction of maximum independent set (MIS) using an adiabatic algorithm, one might apply QSC to identify parts of the graph that have left the satisfying subspace. One then needs to recover into $\mathcal V$ with a particular ansatz -- whether exactly known or computed, else determined by other methods -- that maximizes the probability  algorithm success to the largest degree possible. This recovery is always specifically dependent on the intent of the algorithm.

\section{Exact Distribution Preparation}\label{sec: Exact Dist}

\begin{figure}
    \centering
    \includegraphics[width=2.7 in]{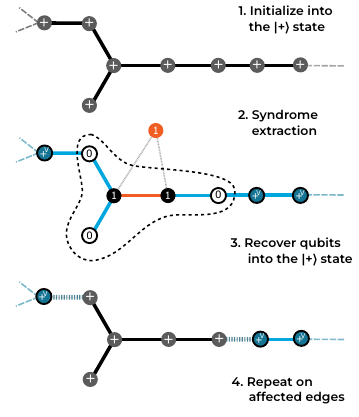}
    \caption{Syndrome extraction depiction with a violation and the recovery sequence. (1) A register is prepared in the $|+\rangle^{\otimes N}$ state. (2) A layer of toffoli gates is applied to the edges of a graph. On read-out of the ancilla, the state is projected into $\bar{\mathcal V}_e$, while the vertices not dominated by the violation locations (identified as blue with $+^{\mathcal{V}}$ label) are in the $|\pis\rangle_g$ state for the violation free subgraph $g$. The syndrome graph, i.e. the location of violations, is constructed. (3) Wherever a violation is found, both the vertices directly adjacent to edge violation and their neighbors must be recovered. If two violations are within distance-1, they can be merged into a larger cluster.(4) The algorithm is then repeated on all edges with an unknown state (black and dashed edges)}
    \label{fig:alg}
\end{figure}
Preparing, manipulating, and extracting information from distributions is computationally expensive for many graph-based problems, making these tasks an important focus in QC applications research. In this section we demonstrate that the existence of a stabilizer on the Independent Set subspace immediately provides a simple subspace correcting algorithm that allows one to exactly prepare probability distributions (or, partition functions) over independent sets of a graph $\mathcal G$. First we will look at the case for preparing a state $|\pis\rangle_{\mathcal G}$ that encodes a perfectly uniform distribution over all IS of a graph $\mathcal G = (V, E)$. The algorithm immediately extends  to the case of preparing Gibbs distributions, denoted by $|\small\lambda_{+^{\small \mathcal V}}\rangle_{\mathcal G}$.

Furthermore, we will identify that this algorithm is the exact quantum analogue of perfect partial rejection sampling. A relatively new algorithm framework first developed in 2016, partial rejection sampling provably produces samples from exact distributions using a stopping condition \cite{guo2019uniform, jerrum2021fundamentals}. The quantum-classical correspondence of this algorithm allows us to make strong statements about run times, as-well as classically simulate run times on any class of graphs up to large sizes, a luxury in quantum algorithm development, and a useful tool for application planning.

\subsection{Algorithm Description (Uniform Distribution)}

In this section we outline a novel quantum approach to constructing the uniform distribution over all independent sets of a graph using only QSC and Hadamard gates. The algorithm is visually demonstrated in Fig. \ref{fig:alg}

For a given graph $\mathcal G = (V, E)$ we assume classical access to an immutable list of edges (E) and vertices (V). We assume a primary quantum register containing $N = |V|$ qubits $\{q_i\}$, whose state is represented by $|\psi\rangle$. WLOG we also assume an ancilla register with $N_A = |E|$ qubits $\{a_e\}$, represented by state $|\phi\rangle$ . 

We require a mutable list of tuples $\text{A}$ to track the edges containing violations and a list of integers $\text{B}$ to track the vertices to be corrected. 

The algorithm is then given here by Algorithm \ref{alg:UDist}:
\begin{algorithm}[H]
\caption{Uniform Distribution Preparation for Independent Set}\label{alg:UDist}
\begin{algorithmic}[1]
\State $|\psi\rangle \gets |\mathbf{+}\rangle$ \Comment{initialize quantum register}
\State  $A \gets$ E \Comment{initialize list with all edges}
\State  $B \gets$ V \Comment{initialize list with all vertices}
\While {len$(A) > 0$}
        \State $|\phi\rangle \gets |\mathbf{0}\rangle$  \Comment{reset ancillae}
        \For{$e \in A$}
            \State Toff$(q_{e[0]}, q_{e[1]}, a_e)$ \Comment{extract syndrome}
        \EndFor
        \State A $\gets$ List[$e$ for $e \in A$ if Measure($a_e$) is 1]
        \State B $\gets$ List[$q_v$ for $v$ in $A$ $\bigcup$ Neighbor($A$)]
        \For{$q_v \in B$}
            \State $|q_v\rangle \gets |+\rangle$ 
        \EndFor
\EndWhile
\end{algorithmic}
Returns: $|\psi\rangle$ in the state $|\pis\rangle_{\mathcal G}$
\end{algorithm}

\subsection{Final State Guarantees}

Algorithm \ref{alg:UDist} always results in a perfect uniform distribution over independent sets, a fact we provide a formal proof of  (and its extension to Gibbs distributions) in Appendix \ref{sec:Proof}. In fact, the intermediate state is always a uniform distribution at the completion of each step. We imagine starting our quantum register in the Hadamard basis $|\boldsymbol{+}\rangle = |+\rangle^{\otimes N}$ and the ancillas in the $|\mathbf{0}\rangle = |0\rangle^{\otimes N_a}$ state. In the bitstring basis, this is represented as:
\begin{align}
| \boldsymbol{+}\rangle|\mathbf{0}\rangle_a = \small\frac{1}{2^{N/2}}\sum_{\textbf{k}=0}^{2^N} |\textbf{k}\rangle|\mathbf{0}\rangle_a
\end{align}

The stabilizing operators are implemented by controlling a set of ancilla-targeting Toffolis on unknown edge states. First consider a graph with $N$ vertices and a single edge, this results in a state entangled across the registers:
\begin{align}
    \text{Toff}_{S_e} | \boldsymbol{+}\rangle|\mathbf{0}\rangle_a  = \small\frac{1}{2^{N/2}}\left(\sum_{\textbf{k} \in  \mathcal V_1} |\textbf{k}\rangle|0\rangle_a +  \sum_{\textbf{k} \in \bar {\mathcal V}_1} |\textbf{k}\rangle|1\rangle_a \right)
\end{align}
Both the left and right sums are uniform superpositions over  a set of bitstrings, with the left (right) being a uniform distribution of all states satisfying (violating) independence on the edge.  Measuring the ancilla qubit applies projective measurements $\Pi_{0} = |0\rangle\langle0|$ and $\Pi_{1} = 
|1\rangle \langle 1|$. Hence, the state is always collapsed into a uniform distribution.

In the case that a violation is found on the edge, we have collapsed the state into:

\begin{equation}
    \sum_{\textbf{k} \in \bar {\mathcal V}_1} |\textbf{k}\rangle|1\rangle_a = |+^{\small \mathcal V}\rangle_{g}|1\rangle_e |1\rangle_a 
\end{equation}
Where $|+^{\small \mathcal V}\rangle_g$ is a uniform superposition of IS on the subgraph $g = G / (A \cup \partial_A)$ which is G with the vertices of the violating edge and the neighboring vertices ($\partial_A$) removed and $|1\rangle_e = |11\rangle_{ij}$ for the vertices $(i,j) = e$ . 

The ancilla and violating qubits are returned to the single qubit $|0\rangle$ states and a Hadamard is applied to the violating qubits, preparing the state $|+^{\small \mathcal V}\rangle_{g} |+\rangle_e |0\rangle_a$, where the procedure is repeated on the edge until success preparing the state $|+^{\small \mathcal{V}}\rangle_{\mathcal G} $ on the primary register.

For the more general case of $|E|$ edges, a similar analysis shows that the distribution is always uniform. In this case the distributions conditioned on the state of the ancilla register measurement still remain uniform. In the event syndrome extraction projects into a state with violations on a set of edges provided by list $A$, in the primary register, one obtains the state $|\psi\rangle =|\pis\rangle_{g} |1\rangle_{A} |\mathbf{0}\rangle_{\partial_A} $
where $|\mathbf{0}\rangle_{\partial_A}$ are all vertices neightbors to vertices in $A$. The form of this result is important for later discussions.

\subsubsection{Extension to Gibbs Distributions}

The classical version of this algorithm has a natural extension to Gibbs distributions. This is obtained by replacing the even sampling of each vertex from the $\{0,1\}$, with drawing from a $\text{Bernoulli}(\tfrac{\lambda}{\lambda+1})$ Distribution, where $\lambda$ is the so-called hardness parameter\footnote{In physics inspired descriptions, this paramter is related to chemical potential $\mu$ and temperature $T$ as $\ln(\lambda) = \mu/T$}. It can be shown that the final probability distribution from this process assigns a weight, up to an overall normalization, $p(s) \propto \lambda^{|s|}$ to a choice of bitstring from the set of independent sets, $s \in IS$, where $|s|$ is the cardinality of the set.

The extension of Algorithm \ref{alg:UDist} to prepare Gibbs distributions over IS similarly requires replacing the Hadamard gates, $H$, which prepare single qubits states $|+\rangle$ with rotated gates, $H_\lambda$ ,that prepare the single qubit states given by:
\begin{equation}
|\lambda_+\rangle = \sqrt{\frac{\lambda}{1+\lambda}}|0\rangle + \sqrt{\frac{1}{1+\lambda}}|1\rangle    
\end{equation}

This state is equivalent to $|+\rangle$ when $\lambda=1$.  The resulting state prepared is given by:
\begin{equation}\label{gibbs}
    |\lambda_\pis\rangle \langle\lambda_\pis|_{\mathcal G} =\frac{1}{Z_{\mathcal G,0}(\lambda)} \sum_{s\in IS} \lambda^{|s|} |s\rangle\langle s|
\end{equation}
where the normalization factor is:
\begin{equation}
    Z_{\mathcal G,0}(\lambda)= \sum_{s\in IS} \lambda^{|s|}
\end{equation}
A proof of this claim is also provided in Appendix \ref{sec:Proof}. The Algorithm \ref{alg:UDist} is then generalized to the following Algorithm \ref{alg:LDist}:
\begin{algorithm}[H]
\caption{Gibbs $\lambda$ Distribution Preparation for Independent Set}\label{alg:LDist}
\begin{algorithmic}
\State $|\psi\rangle \gets |\mathbf{\lambda}\rangle$ \Comment{initialize quantum register}
\State  $A \gets$ E \Comment{initialize list with all edges}
\State  $B \gets$ V \Comment{initialize list with all vertices}
\While {len$(A) > 0$}
        \State $|\phi\rangle \gets |\mathbf{0}\rangle$  \Comment{reset ancillae}
        \For{$e \in A$}
            \State Toff$(q_{e[0]}, q_{e[1]}, a_e)$ \Comment{extract syndrome}
        \EndFor
        \State A $\gets$ List[$e$ for $e \in A$ if Measure($a_e$) is 1]
        \State B $\gets$ List[$q_v$ for $v$ in $A$ $\bigcup$ Neighbor($A$)]
        \For{$q_v \in B$}
            \State $|q_v\rangle  \gets |\lambda_{+}\rangle$ 
        \EndFor
\EndWhile
\end{algorithmic}
Returns: $|\psi\rangle$ in the state $|\lambda_\pis\rangle_{\mathcal G}$
\end{algorithm}

\subsection{Runtime analyses}

The runtime of Algorithm \ref{alg:UDist} can be carefully bounded analytically, or simulated classically without the need for quantum simulation techniques, using the exact classical mapping to partial rejection sampling of the previous section. To consider the runtime, we generalize to the case of the Gibbs distribution in hardness-parameter $\lambda$, where $\lambda = 1$ corresponds to the uniform distribution case. We can then directly quote the results of runtime analyses from the classical process.

For graphs of bounded degree-$d$, \cite{guo2019uniform} showed, analytically, that hardness parameters bounded by the inequality
$$
\lambda < \frac{1}{2\sqrt{e}d -1}
$$
had a worst-case runtime that was, on average $O(n)$, and with a high probability of $O(n\log n)$. This is a lower-bound on $\lambda^*_d$, which represents the largest $\lambda$ for which all graphs of bounded degree $d$ converge with this expected runtime. It asymptotically coalesces with the bound $\bar \lambda_d^* \sim O(d^{-1})$ for which it becomes provably NP-hard to sample from this class of graphs \cite{sly2012computational, weitz2006counting, galanis2016inapproximability}. 

The runtime estimate for the classical algorithm was tightened for $d=3$ in Ref. \cite{jerrum2021fundamentals} to $\lambda^*_3 \geq $0.150, where slack still exists in the analysis. The strict upper-bound on this quantity is given by $\lambda_3^* < 0.5-\epsilon$. For this reason, the preparation of uniform states $(\lambda = 1)$ will in such classes of graphs in $d=3$ will always be exponential for the average case. When looking at the preparation of Gibbs states on $3$-regular graphs, we find that for the average case one obtains sublinear preparations for $\lambda \lesssim 0.7$

With the knowledge of these analytical bounds, in the next section we explore numerically some graph classes up to size $n = 80$. We do this by directly applying the classical algorithm for partial rejection sampling to graphs chosen from random using either an appropriate random sampling function from the \verb|networkX| package when available, or through careful construction of a unbiased graph sampling algorithm. Mean and median data is calculated from at least 100 samples for each graph size point, with all collected data plotted in a semi-transparent color.  In particular, we look at scaling for planar graphs, sub-graph approximations, and then finally look at empirical results for Gibbs state preparation, where we find sub-linear scaling on average for larger $\lambda$ values than the provided bounds on average case instances.

\subsubsection{Empirical runtime for uniform preparation on regular planar graphs}

\begin{figure}[t]
    \centering
    \includegraphics[width=3.4in]{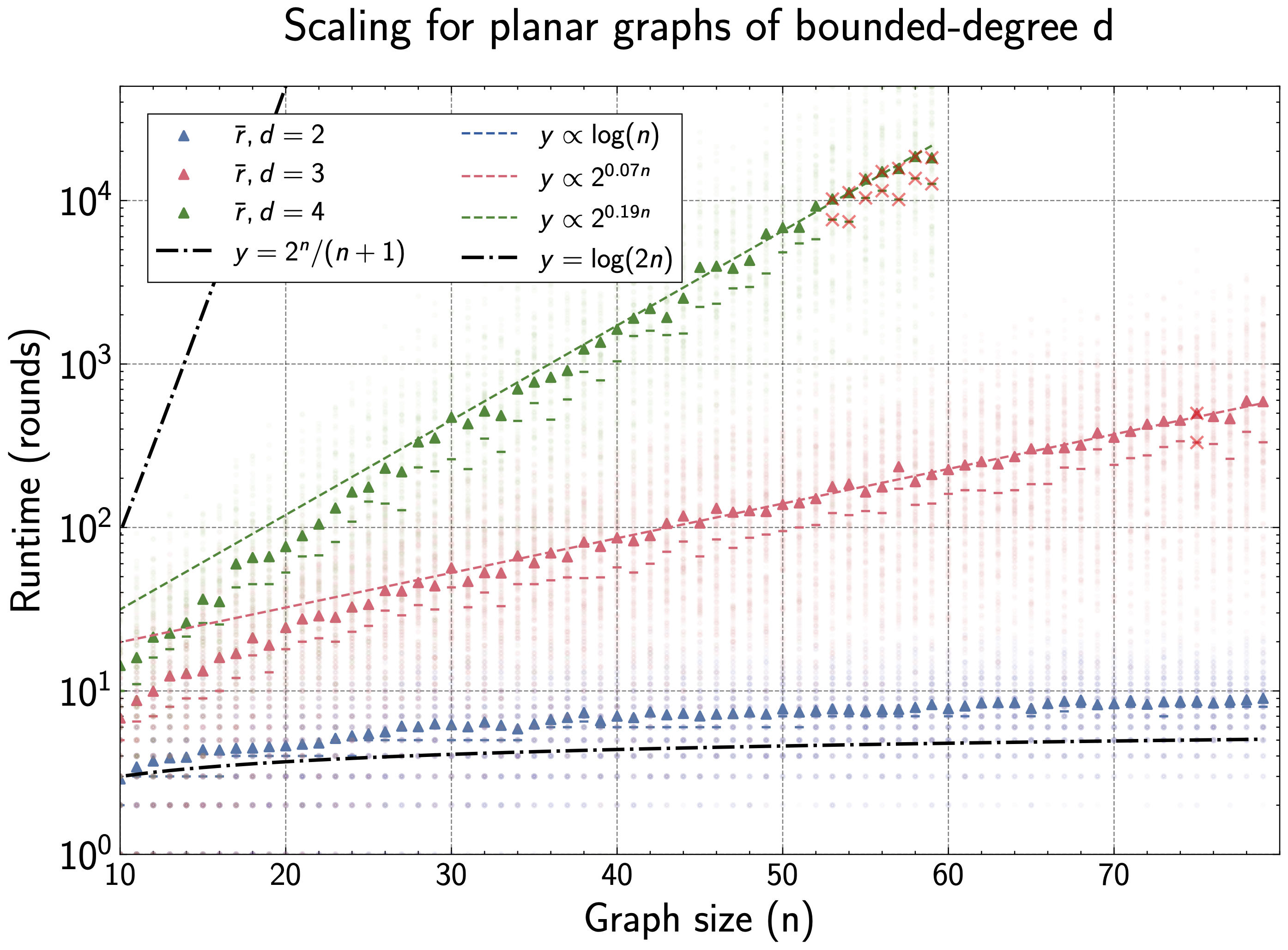}
    \caption{Plots of uniform preparation runtime vs. graph size in planar graphs of bounded-degree $d$, obtained by running the classical dual algorithm up to graph size $n=80$.  Triangular data points represent mean runtimes for 200 random graphs for each $n$. The dash marker represents the median. Each semi-transparent point on the plot represents a single run. Red `x' marks indicate that the runtime exceeded the maximum number of tried ($5\times 10^{5}$ for $d=4$ and $5 \times 10^4$ for $d=3$). For planar graphs of size $d=3$ we observe weakly exponential runtimes asymptotically for the average case. The dashed lines represent the best and worst possible runtimes for any graph of size $n$. }
    \label{fig:bounded_planar}
\end{figure}

The cited papers on partial rejection sampling in independent sets provided extensive analytical and simulation-based evidence for runtime bounds in graphs of bounded degree-$d$. Because planar graphs imbue additional structure, we considered that runtimes might be improved in this class of bounded degree graphs. 

As shown in the $\log-\log$ plot in Fig. \ref{fig:bounded_planar}, we found that the average runtime for planar graphs of bounded degree $d=3$ asymptotically limit to weak exponential scaling to the best of our fits. The plots were made by running the algorithm for $200$ random graphs per data point and recording the number of rounds required to reach the algorithm termination condition. The random graphs were drawn from a custom script that uniformly samples planar graphs of a given bounded degree. The histogram of runtimes is heavily skewed toward fewer rounds, as indicated visually \footnote{Note that this is a lin-log plot, so the approximately normal sample appearance about the line of fit corresponds to a skewed log-normal distribution in the raw data} and by the median (dashed points), with a long tail of low probability-high rounds events. Planar graphs of bounded degree $d=4$ are clearly exponential beyond a graph size of $n=30$, and $d=2$ has clear logarithmic scaling in graph size.

Despite weakly exponential behavior, the number of QSC rounds for graph sizes required for many real-life applications (as few as $O(100)$) would not be prohibitive to using the algorithm to prepare states for further processing if a uniform distribution was required for the application. It is also important to remember that the number of rounds required is not the number of repeated circuit executions, but is instead proportional to the total (logical) circuit depth.

\begin{figure*}[t]
    \centering
    \includegraphics[width = 3.3 in]{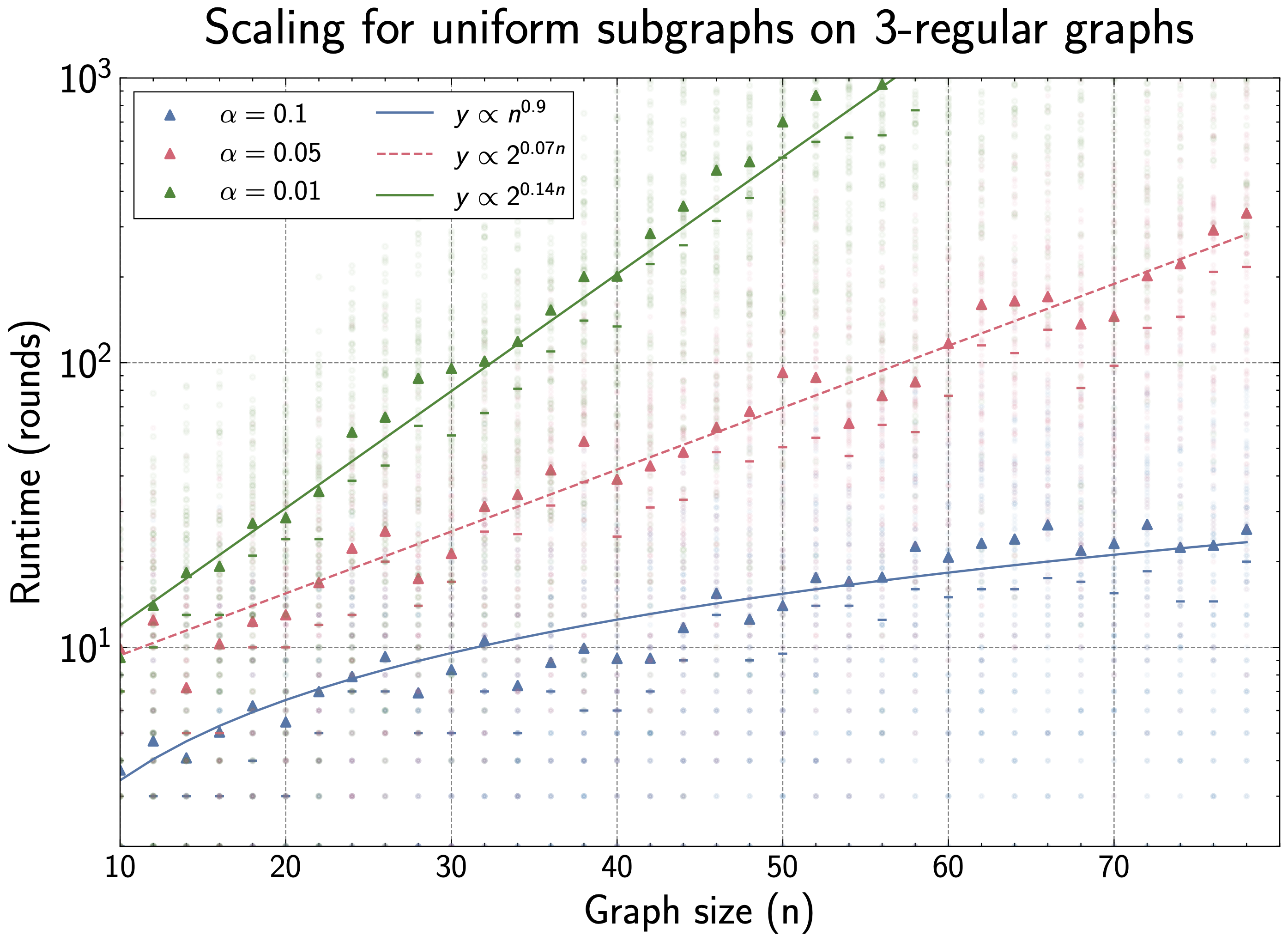}
    \hfill
    \includegraphics[width=3.3 in]{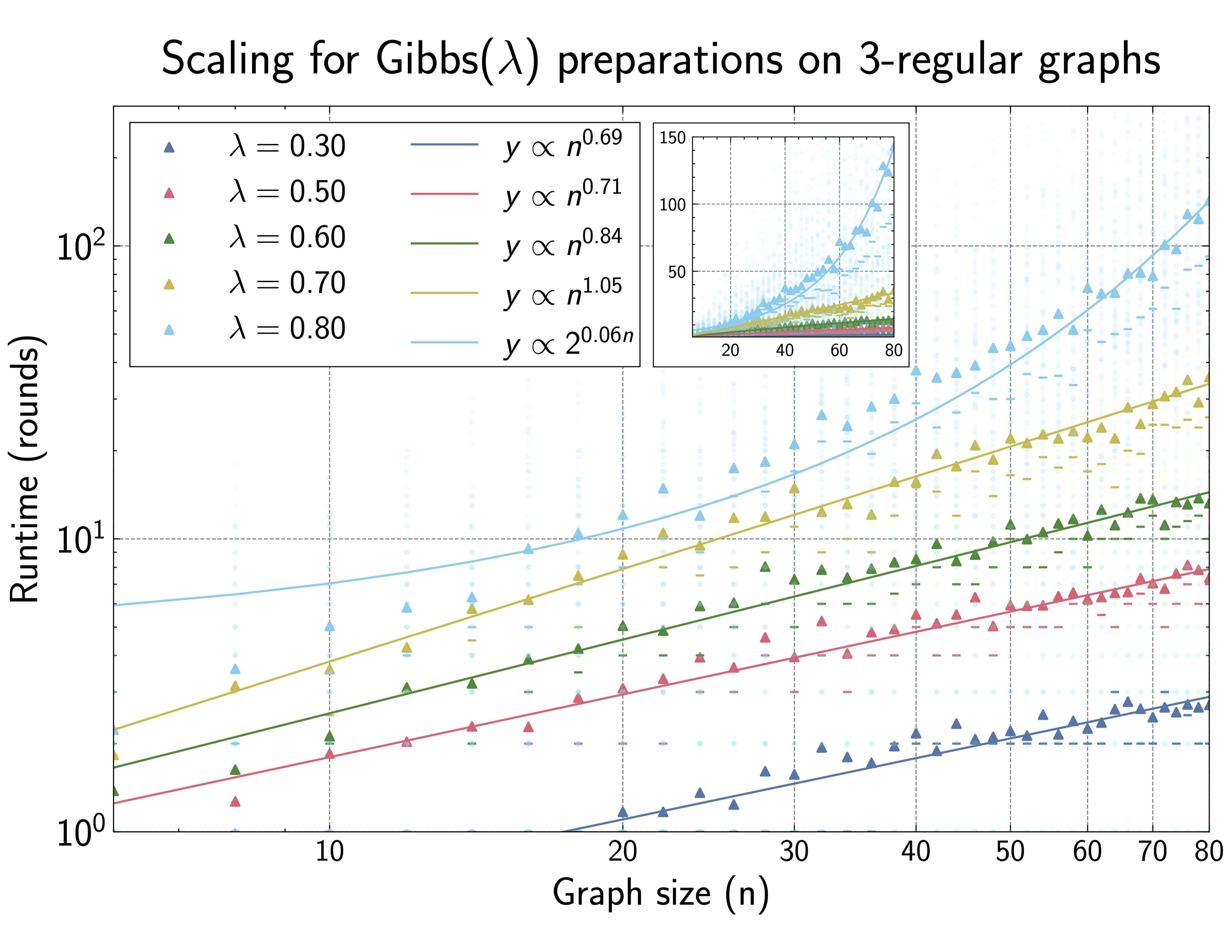}
    \caption{Triangular data points represent mean runtimes for 100 random graphs for each $n$. The dash marker represents the median. Each semi-transparent point on the plot represents a single run. (Left) Scaling for preparation of $g_\alpha$- subgraphs. Our fits reveal that, at least for graph sizes up to $n\sim 80$, halting states for $\alpha \geq 0.10$ have runtimes that scale efficiently with graph size. (Right) Gibbs preparation of various $\lambda$ states on $3-$regular graphs. While the loose analytical bound for efficient scaling for worst-case graphs are around $\lambda^*_3 \sim 0.15$, we find that, empirically, average case graphs scale sublinearly in rounds up to $\lambda \approx 0.7$. The inset provides the linear-linear view of the plot to show the phase-transition in hardness more obviously.}
    \label{fig:othercases}
\end{figure*}

\subsubsection{Halting states on $\alpha-$subgraphs }
Rather than run the algorithm to completion, one may alternatively choose an early halting condition based on the fraction of edges, $\alpha = |A|/|E|$, that are permitted  to have violations. This results in a termination into the state:

\begin{equation}
    |\psi_\alpha\rangle = |\pis\rangle_{g_{\alpha}} |1\rangle_{A} |\mathbf{0}\rangle_{\partial_A}
\end{equation}
where $g_{\alpha} = G \backslash A \bigcup \partial_A$ is the subgraph less the vertices associated with violating edges. 

Similar to before, each data point represents runs from $100$ random 3-regular graphs generated using the \verb|networkX.random_regular_graph()| function for all allowable graph sizes up to $n=80$ for $d=3$ . From our numerical evidence in Fig \ref{fig:othercases}, it seems that for graph sizes up to $n\approx 80$, approximations  with $\alpha \approx 0.1$ can be prepared sub-linearly in the average case. These results should be continued out to larger graph size using a more streamlined simulation technique, or verified analytically.

From such a state, one can apply a unitary operation on $|1\rangle_{A} |\mathbf{0}\rangle_{\partial}$ to prepare a target uniform state which spans a larger fraction of the states. For example, one might prepare $|\pis\rangle_e$ sequentially on each violating edge $e$ to obtain a uniform distribution on the subgraph induced by $G \backslash B$, where the boundary $|0\rangle_\partial$ vertex states are collapsed.  This, in practice on graphs small enough to simulate quantum circuits on ($n\le 15$),as results in uniform distributions that span a reasonable subset of the IS. Due to how small the graphs are, and the likelihood of making analytical progress on this question in an upcoming publication, we do not include these small numerical studies here.

In principle, one is free to construct a clever oracle involving a nontrivial classical computation of a unitary $U_{repair}$ which is capable of transforming violation and boundary regions into an acceptable state for the intended application.

\subsubsection{Sublinear preparation of Gibbs distributions}

As mentioned, the algorithm can be adapted to prepare Gibbs distributions over the independent sets, as states of the form of Eq. \eqref{gibbs} , rather than uniform distributions. These distributions fully span the independent sets, but have a weight dependent on the chosen hardness parameter $\lambda$.

We numerically evaluated the expected runtime for preparing Gibbs distributions on $3-$regular graphs for various $\lambda$ in Fig. \ref{fig:othercases}. Each data point in the plot corresponds to $100$ random graphs generated by the \verb|networkX.random_regular_graph()| function for all allowable graph sizes up to $n=80$ for $d=3$. 

While the limits worked out analytically refer to average runtimes of worst-case graphs in their respective classes, we find that up to about $\lambda \sim 0.7$ we are able to prepare distributions for random graphs efficiently, on average.

This result is quite remarkable when we consider the value of perfectly prepared q-sampling state when used as an input to other algorithms such as Grover's Search, Quantum Counting, or potential distribution comparison algorithms\cite{grover1996fast,brassard1998quantum,bravyi2011quantum}. We remark on future work towards this goal in the discussion of Section \ref{sec:discuss}.

\subsection{Discussion}\label{sec:discuss}

The mapping to classical runtimes is a double-edged sword as, for the case of \textit{direct} sampling from perfect distributions, this algorithm can not produce a single sample more efficiently than a classical computer. We note that the output of this algorithm, however, is the \textit{entire distribution} amplitude encoded into the quantum register, while the output of the classical sampling algorithm is a single bitstring. We note that some algorithms for approximately uniform sampling have been previously devised using adiabatic\cite{PhysRevA.101.012318} and Markov methods\cite{wild2021quantum}, although these are approximate preparations. 

We argue that the ability to produce \textit{perfect} distributions, $|\psi\rangle = \sum_{s \in \textrm{IS}} \sqrt{p_s} |s\rangle$, with known distribution properties can be considered a resource, especially when these distributions can be prepared in sub-linear time like the Gibbs distributions for sub-critical $\lambda$. Such a resource state, which spans the entire basis of independent sets, may serve as the first step in an algorithm that probes quantitative proprieties of the graph. For example, using amplitude amplification\cite{brassard2000quantum, grover1996fast} , one could find a set of independence number $k$ within $O(N\small\sqrt{Z_{G,0}(\lambda)/m_k \lambda^k})$ time, where $m_k$ is the multiplicity of sets of independence $k$, for $\lambda < \lambda^*$. One may also use such states as the inputs to distribution tests, for example using orthogonality tests as in Ref \cite{bravyi2011quantum}.

Another interesting property of the distribution algorithms is that if the phases of the states in the distribution do not matter for the follow up to the algorithm, then one only needs to correct for bit-flip errors during the state preparation. The result from a bit-flip corrected implementation of this code will still have a classical probability distribution encoded into the amplitude of states, but the phase from those states will be randomly distributed according to the mechanism of phase errors. This code is also robust to leakage errors due to qubit loss, since detection of leakage and atom replacement can be built into the QSC protocol.  

\section{Applications to Constrained Optimization: Adiabatic Depth Reduction}\label{sec: adiabatic}
In addition to the distribution preparation protocol, one can use QSC as a way to stabilize constraint subspaces during the execution of an optimal state preparation algorithm. The typical way constrained optimization problems are distilled for quantum computation is to map them to an Ising-like Hamiltonian, most famously demonstrated in Lucas' \textit{Ising forms of many NP Problems}\cite{lucas2014ising}. The Hamiltonian of these problems have a common structure, namely, they are typically the sum of two component Hamiltonians:

\begin{equation}\label{optimization}
    H = \lambda H_A + H_B,
\end{equation}
where $H_A$ encodes a set of \textit{constraints} on the qubits as its lowest energy state, and $H_B$ encodes the \textit{objective} as its lowest energy states, and $\lambda$ is some Lagrange multiplier that roughly signifies the importance of the constraints in the energy landscape. An adiabatic, or adiabatic-inspired algorithm is then run to find the low-energy states of this problem.

An obstacle for such approaches is that, often, such algorithms return invalid --- \textit{i.e.} constraint violating states --- with high probability. The reason for this is either due to expressibility of parameterized variational ansatz, such as in QAOA, or due to errors from non-adibatic terms in adiabatic evolution or the Trotterization of it. Given this, it is reasonable to ask if QSC can improve adiabatic algorithms by explicitly constraining it to run in the intended subspace.  A related study in Ref \cite{herman2023constrained}  used the quantum Zeno effect to monitor constraints without recovery, in tandem variational algorithms such as QAOA, and found evidence for solution quality improvement in very small problem sizes. As we are not interested in variation algorithms, and rather coherent algorithms of the form of shown in Fig. \ref{fig:NAQC capabilities}, we instead look at how QSC could help reduce the depth needed for optimization problems using adiabatic preparation.

In this section, we demonstrate an example of this idea using QSC to reduce the circuit depth needed to approximate solutions to Maximum Independent Set (MIS) for a given graph $\mathcal G = (V, E)$. We provide details for the full algorithm,  including the evolution unitary and a recovery strategy. We are only able to simulate this method on small graph sizes due to the computational intensity of adaptive circuits.

\subsection{Non-Abelian Adiabatic Mixing }
First we describe the Non-Abelian adiabatic evolution algorithm for preparing MIS.  In the notation of \eqref{optimization}, the problem of MIS is given by the constraint Hamiltonian:

\begin{equation} \label{eq:MIS_constraint}
    H_A = -\Delta \sum_{\langle ij \rangle} Z_i + Z_j - Z_iZ_j
\end{equation}
where the sum runs over all edges $\langle ij\rangle $ in $\mathcal G$, $Z_i$ are Pauli-Z matrices acting on the $1-$qubit subspace of the $i$th, and $\Delta$ is an energy scale. The objective Hamiltonian is given by:

\begin{equation}
    H_B = - \sum_{i} Z_i
\end{equation}
 such that maximizing the number of qubits in the single-qubit $|1\rangle$ state minimizes the value of this term. 

 We utilize a non-abelian adiabatic state preparation algorithm, $U_A(t)$, first detailed by Yu, Wu, and Wilczek \cite{yu2021quantum, PhysRevA.101.012318}. We chose this algorithm due to the fact that there is a large gap protecting the independent set subspace at all points in the evolution. This method of state preparation differs from standard annealing which relies on the ramping-out of a transverse field. Rather, the constraint Hamiltonian $H_A$ is taken to be the problem Hamiltonian, and the optimization is carried out by a slow global rotation applied the quantum register. Prior to the start of the algorithm, the quantum register is prepared in a \textit{least optimal} state --- \textit{i.e.} a state that maximizes $H_B$ ---  within $\mathcal V$ is prepared. In the case of maximum independent set, this is simply $|\mathbf{0}\rangle$. The combined action of $H_A$ and the slow global rotation induces mixing in the ground-state subspace only, evolving the register toward the most optimal state while staying withing $\mathcal V$ for a sufficiently slow rotation. 
 
 One should note that the form of Eq. \eqref{eq:MIS_constraint} is, up to a rescaling and shift, identical to the stabilizer for independent set. The sum of this Hamiltonian runs over edges, and on each edge, the Hamiltonian has the spectrum:

 \begin{equation}
    E_{\langle ij \rangle} = \begin{cases}
  -\Delta  & |00\rangle_{ij},\; |01\rangle_{ij}, \; |10\rangle_{ij} \\
  3\Delta\ & |11\rangle_{ij}
\end{cases},
\end{equation}

The Hamiltonian contains a $\Delta E =4\Delta$ gap between the ground state and the first excited state. The ground-state subspace of this Hamiltonian, and hence the $\mathcal V$ of the full optimization problem, is the set of independent sets of a graph $\mathcal G$.

We use the same complex representation of the global  $\text{SO}(3)$ rotation\footnote{The sign differences in our rotation matrix are due to identifying the $|1\rangle$ state as set inclusion, while the previous authors considered $|0\rangle$ to indicate inclusion.} over $N$ qubits presented in the original paper, namely:

\begin{equation}
    U_B(\theta,\varphi) = \begin{pmatrix}
        -\cos(\theta/2) & e^{i\varphi}\sin(\theta/2) \\
         e^{-i\varphi}\sin(\theta/2) & \cos(\theta/2)  \\
    \end{pmatrix} ^{\otimes N},
\end{equation}
where $\theta = \theta(t)$ and $\varphi = \varphi(t)$ are parameterized functions of time.  The ideal adiabatic evolution then takes the form

\begin{equation} \label{eq:UA}
    U_A(t) |\psi(0)\rangle = e^{ - i \int _0^{t}  H(t')dt'  }|\psi(0)\rangle= |\psi(t)\rangle
\end{equation}
for $H(t) = U_B(t) H U_B^{-1}(t)$. 

At the end of the evolution, a bit flip operation is applied to all qubits in the register, and the register is read out in the computational basis. If the evolution was sufficiently slow enough, \textit{e.g.} slow enough to not close the exponentially small gap in the effective gauge Hamiltonian\footnote{This is different than the gap in the constraint Hamiltonian, and is instead is the gap separating states respresenting sets of maximum independence }, then the measured state will, with a very high probability, be a maximum independent set.
The authors of Ref \cite{yu2021quantum} showed that for parameters $T = N^2$, $\dot \theta = \pi t/T$, $\dot \varphi = t$, where $T$ is the total runtime, the algorithm generally produces MIS with a high probability.

\subsubsection{Gate-based Algorithm}

To target a universal gate-based platform, we must Trotterize this unitary evolution and convert it to gates, which introduces additional errors that lead to the quantum register to returning invalid states when measured in the computational basis. We  use a first-order Trotterization to convert $U_A(t)$ into a form appropriate to transpile into a gate sequence:

\begin{equation}\label{eq:trotter}
\begin{aligned}
U_A(t) &\approx \prod_{n=0}^{N_T} U_B(n\, dt) e^{ - i dt H_A} U_B^{-1}(n\, dt) \\
&\approx  U_B(N_t)\prod_{n=1}^{N_T}  e^{ - i dt H_A} U_B^{-1}(n\, dt) U_B\left((n-1)\, dt\right)\\
& \ = U_B(N_t) \prod_{n=1}^{N_T} \delta U_A(n)
\end{aligned}
\end{equation}
\begin{figure}[t]
    \centering
    \includegraphics[width = 3in]{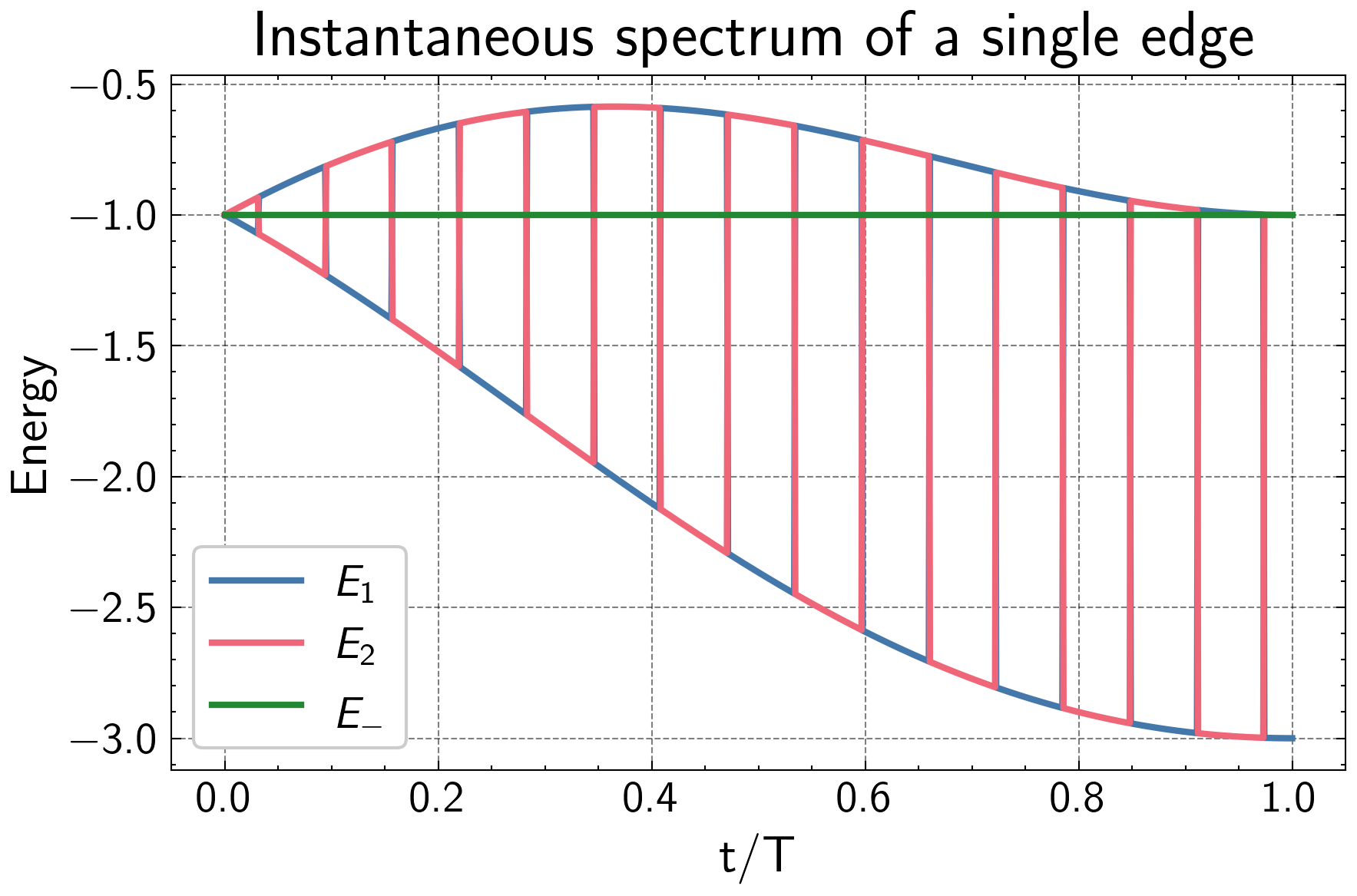}
    \caption{Exact instantaneous spectrum of the Non-Abelian Adiabatic algorithm for MIS, calculated for a single edge. The $|-\rangle$ Bell State always has constant energy in the moving frame. The ground state of the system is a time-dependent function of the remaining two constraint-satisfying basis states. The frequency of oscillation between the two states is dependent on $\dot \varphi$.}
    \label{fig:energies}
\end{figure}
where in the second line we have shuffled the terms so that after every application of $\delta U_A(n)$, the quantum register is in the basis that the constraint $H_A$ is applied in. The leftover $U_B(t)$ is a full $\theta=\pi$ rotation of the register and can be conveniently omitted to prevent having to classically apply bit-flips to each element of the bitstring as is done in \cite{yu2021quantum}.

Using parameter selections guided by Ref.\cite{yu2021quantum}, -- hence restricting to a good approximation of adiabatic evolution in continuous time -- Trotter step size becomes the free parameter in this algorithm that controls depth. Trotter errors should play a dominant role in the creation of excitations out of the ground state manifold,  unless the step size chosen is below the Trotter transition. This transition has been identified by multiple authors  \cite{sieberer2019digital,heyl2019quantum} and is reproduced empirically here. 

\subsection{QSC for Non-Abelian MIS}
\begin{figure*}
\centering
    \includegraphics[width = 3.3 in]{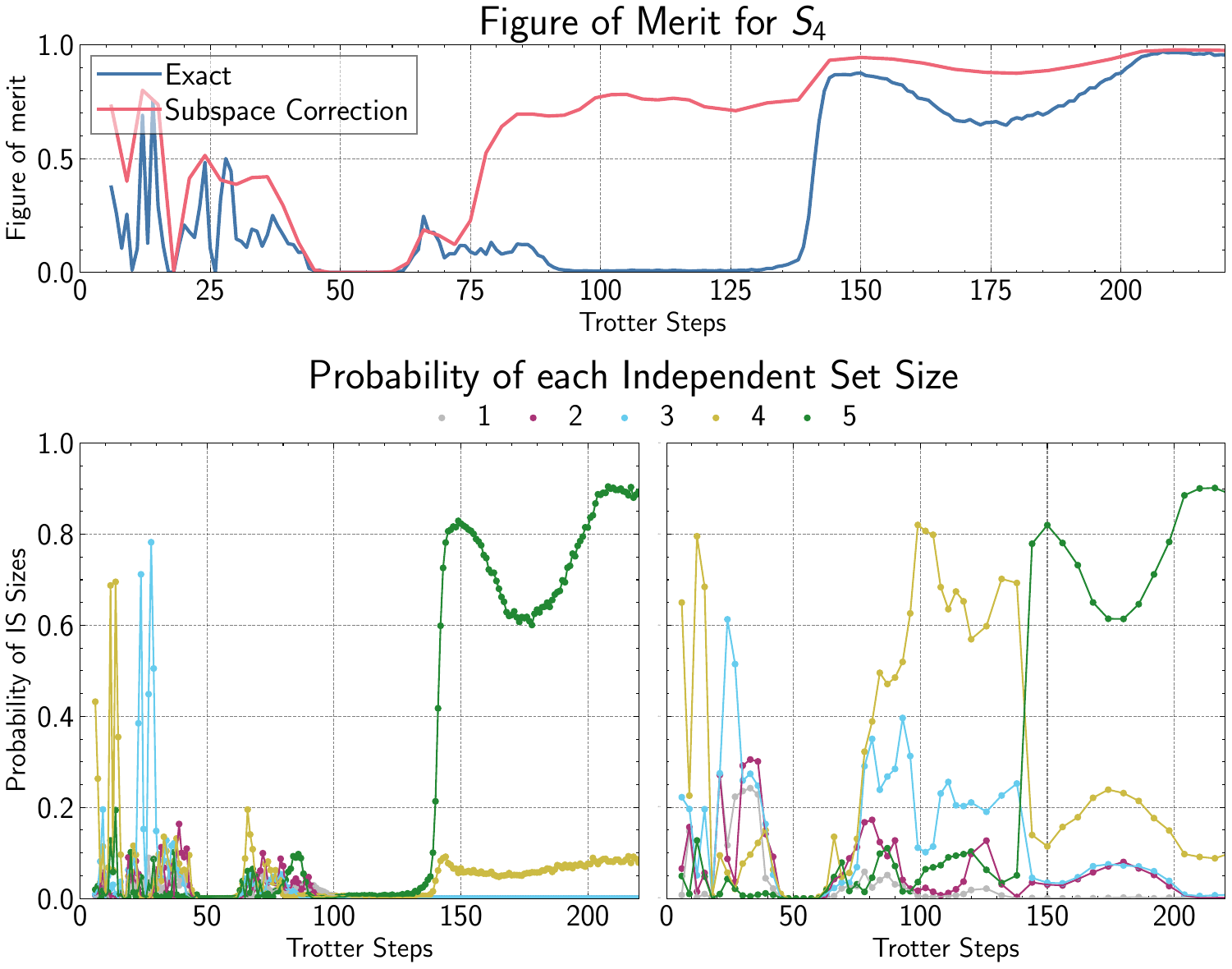}
    \hfill
    \includegraphics[width = 3.3 in]{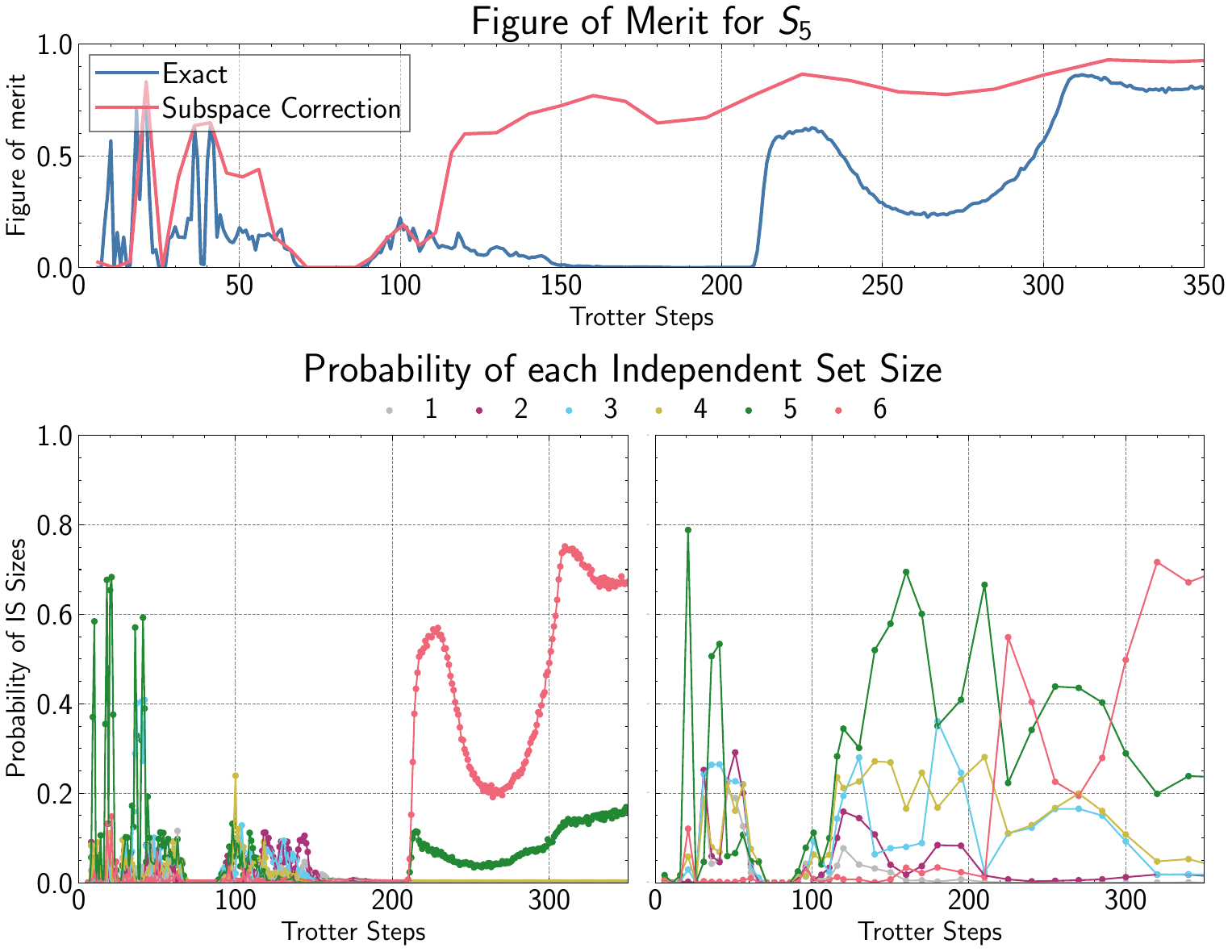}
    \caption{Metrics for solution quality in graphs $S_4$ (left) and $S_5$ (right) compared between the bare non-Abelian adiabatic algorithm and the same algorithm using QSC with the isolated edge recovery applied at regular intervals. In this simulation, we evolve to time $T= N^2$ with increasing numbers of Trotter steps. The top plots of both images show the figures of merit, which is the average size of the independent sets over the size of the maximum independent set. The bottom plots compare the probability of finding each independent set size for the exact algorithm (left) and the algorithm with QSC (right). The figure of merit plots show that solution quality dramatically improves in in much fewer Trotter steps as compared to the exact solution. An inspection of the probability of finding different IS solutions (bottom plots) shows that the figure of merit is boosted by a high probability of finding a great approximation to MIS. Asymptotically the solutions converge to the sane probabilities as the rate of errors vanish through the Trotter transition. }
    \label{fig:ssc-adiabatic}
\end{figure*}
QSC is applied in the Trotterized algorithm of \eqref{eq:trotter} by interleaving the stabilizer extraction and recovery gates between successive $\delta U_A(n)$ -- hence, while the quantum register is in the correct basis to respect constraints. QSC need not be applied every step. An analysis to understand the circuit-depth trade space, similar to when choosing the correct product formula order, should be carried out.

\subsubsection{Recovery Operation}
Up until now, we haven't discussed the use of a tailored recovery operation for QSC. We are now in the context of an optimization problem and must consider an appropriate ansatz based on our evolution unitary.

Because our state preparation is adiabatic, we expect our quantum register to, ideally, remain in the  instantaneous ground-state (or at least the low-enegy manifold) of a time dependent Hamiltonian, which eventually evolves to have MIS as the ground state. 

In the moving frame of a Hamiltonian dependent on parameters $\Phi$, a state evolves as:
\begin{equation}
    i\partial_t |\bar \psi\rangle = \left( U^\dagger H U - i \dot\Phi U^{\dagger}\partial_\Phi U\right) |\bar\psi\rangle
\end{equation}
We define:
\begin{equation}
    \bar A_\Phi = i  U^{\dagger}\partial_\Phi U
\end{equation}
so that in the moving frame:
\begin{equation}
    \bar H = H + \dot\Phi A_\Phi
\end{equation}

$\bar H$ with $\Phi$ and its eigenvectors and eigenvalues can be calculated exactly for a graph with two vertices and a single edge between them after a first-order expansion in slow parameters $(\dot\theta, \dot\varphi)$. In particular, the eigenstates are $|11\rangle$ with an energy $E_{11} = 4\Delta$, the Bell state $|\Phi^-\rangle = \frac{1}{\sqrt 2}(|01\rangle - |10\rangle)$ with an energy $E_-=0$ and a time dependent mixture of the remaining two eigenstates which are combinations of the $\{|00\rangle, |01\rangle, |10\rangle\}$ basis states. A plot of the instantaneous energies of the subspace for the single edge problem is provided in Fig. \ref{fig:energies}.  

When running the ideal algorithm on a single edge,  the exact recovery upon projecting into the violating state $|11\rangle$ should be applying a unitary to prepare the lowest energy state at time $t$. This can be done by numerically computing the form of the lowest energy eigenstate at time $t$ , and applying a gate sequence to prepare that two-qubit state. The general parameterized sequence is provided in \ref{sec:sequence} for applying this in practice.

In a larger graph, this method of determining the correct recovery faces two issues. The first is that finding the optimal target state of a recovery in the same way would require classically simulating the entire quantum system, which is self-defeating. Moreover, once we have a graph with multiple adjacent edges, syndrome extraction of violations result in projecting boundary verticies into the $|0\rangle$
state, similar to the sampling algorithm. This results in an overall degradation of the solution since the objective is to find MIS, or an approximation.

Despite these points, in the next section, we will consider using a simple recovery strategy inspired by the isolated edge case. Empirically, we tested multiple local recovery strategies into the independent set subspace of general graphs under adiabatic evolution, including preparing violating edges into $|00\rangle$, $|\Phi^-\rangle$, and $|+\rangle$. By far, the best results were obtained when we replaced the two qubit state of any detected violation with the wavefunction of the isolated edge system under the same evolution at the appropriate time $t$. This approximation appears to work well in small graph sizes with low connectivity, and leaves room for further improvements. Due to the limitations of the software used for simulation, it was not viable to apply recovery strategies that were functionally dependent on the read-out of multiple ancilla qubits.

\subsection{Numerical Results}

Empirically, we find that that QSC  protects against Trotter errors by exploring how algorithm outcomes behave as a function of Trotter step size with all other algorithm parameters held fixed. To limit computational resources expended in this preliminary study, we constrained ourselves to $S_n$ graphs described in Ref \cite{yu2021quantum} , which feature some nice properties and an exponentially small gap  in the instantaneous frame. We chose to apply QSC once every $10$ Trotter steps, or once every $N_T/4$, whichever is smaller. 

For the recovery, we applied a recovery inspired by the isolated edge wavefunction described, limited by the functionality of commercially available simulators. We performed syndrome extraction in a serial fashion on each edge. If during extraction we detected a violation of the constraint subspace, we reset the two edge qubits and prepared the state that an isolated edge graph would be in under the same evolution protocol. We then move to the next edge in the graph. This algorithm is not perfectly identical to reading out all ancillas and then determining a recovery.

Fig. \ref{fig:ssc-adiabatic} shows our limited numerical results. The behavior of the adiabatic algorithm is complicated even in the exact case. The small region of high probability to find a good approximation at $O(10)$ steps appears to be an artifact that vanishes with increasing problem size and seems to be dependent on the numerical value of $\dot \varphi$, and merits further investigation. We instead focus on the behavior at increasing step sizes, where we see that at about half the depth of the transition in Trotter error for the exact case, we see a transition in solution behavior in the QSC version. In particular, we see that the next best approximation to MIS becomes the most probable bitstring. The two solutions asymptotically converge as Trotter error vanishes.

\subsection{Discussion}\label{sec:discuss2}

The numerical results of this study should be taken with caution as the graphs investigated are very small. Moreover, we explored only a limited class of graphs for this study.  Due to the structure of the $S_n$ graphs, it is possible that the isolated edge ansatz performs unusually well. We did see similar behavior in the handful of very small (up to size 12) $3$-regular graphs simulated, however, we did not do exhaustive testing of random graphs due to the computational expense of simulating this algorithm and have omitted these results at present. In future work, we hope to construct a tailored simulation pipeline to speed-up explorations of random graph classes, and also to explore recovery techniques with more complexity. Unfortunately, feed-forward circuits are heavily resource intensive.

That being said, it is quite surprising how well the isolated edge ansatz performs in practice. This inspires optimism that more detailed recovery strategies that involve harder classical computations and larger recovery clusters could potentially future reduce depths needed for adiabatic algorithms.

\section{Conclusion}

Taking inspiration from error correcting codes, we developed the idea of constructing sets of operators to stabilize constraint subspaces that arise in Ising forms of many computational problems. We provide a recipe for constructing these stabilizers in Appendix \ref{app:detections}. We explicitly explored these stabilizers for the problem of independent set, and demonstrated two techniques to utilize them for constrained problems: distribution preparation and depth reduction. The tools implemented in this work make use of new functionality available on NAQC hardware, and we believe that they will serve as useful building blocks for next generation algorithm development.

Using these tools, we were able to construct a distribution preparation algorithm that can exactly prepare Gibbs distributions over independent sets of a graph when coupled with a stopping conditions. We found that for sub-critical values of the hardness parameter $\lambda$, one may prepare some of these distributions in sub-linear time. We were also able to reduce the depth required to adiabatically prepare a good approximation to maximum independent set using QSC in conjunction with a specialized adiabatic algorithm.

In future work, we plan to provide more a more rigorous analysis for many of the observations in this paper. We are also investigating deeper links between the connections that exist between classical and quantum partial rejection sampling, and possible algorithms to make use of distribution preparation. For QSC, a promising direction for application includes using it to prepare or maintain constrained subspaces that are not difficult to prepare, but are nevertheless embedded in optimization problems of industrial or technological value.

\begin{acknowledgments}
We wish to acknowledge Eliot Kapit for discussion and comments during the early development of this technique. KP would like to acknowledge Tim Hseih for general discussions about adaptive circuits and Eric Jones for prior helpful discourse regarding the non-abeliean adiabatic unitary that was ultimately used as an example in this paper. 

\end{acknowledgments}
\bibliography{apssamp}

\begin{thebibliography}{35}%
\makeatletter
\providecommand \@ifxundefined [1]{%
 \@ifx{#1\undefined}
}%
\providecommand \@ifnum [1]{%
 \ifnum #1\expandafter \@firstoftwo
 \else \expandafter \@secondoftwo
 \fi
}%
\providecommand \@ifx [1]{%
 \ifx #1\expandafter \@firstoftwo
 \else \expandafter \@secondoftwo
 \fi
}%
\providecommand \natexlab [1]{#1}%
\providecommand \enquote  [1]{``#1''}%
\providecommand \bibnamefont  [1]{#1}%
\providecommand \bibfnamefont [1]{#1}%
\providecommand \citenamefont [1]{#1}%
\providecommand \href@noop [0]{\@secondoftwo}%
\providecommand \href [0]{\begingroup \@sanitize@url \@href}%
\providecommand \@href[1]{\@@startlink{#1}\@@href}%
\providecommand \@@href[1]{\endgroup#1\@@endlink}%
\providecommand \@sanitize@url [0]{\catcode `\\12\catcode `\$12\catcode `\&12\catcode `\#12\catcode `\^12\catcode `\_12\catcode `\%12\relax}%
\providecommand \@@startlink[1]{}%
\providecommand \@@endlink[0]{}%
\providecommand \url  [0]{\begingroup\@sanitize@url \@url }%
\providecommand \@url [1]{\endgroup\@href {#1}{\urlprefix }}%
\providecommand \urlprefix  [0]{URL }%
\providecommand \Eprint [0]{\href }%
\providecommand \doibase [0]{https://doi.org/}%
\providecommand \selectlanguage [0]{\@gobble}%
\providecommand \bibinfo  [0]{\@secondoftwo}%
\providecommand \bibfield  [0]{\@secondoftwo}%
\providecommand \translation [1]{[#1]}%
\providecommand \BibitemOpen [0]{}%
\providecommand \bibitemStop [0]{}%
\providecommand \bibitemNoStop [0]{.\EOS\space}%
\providecommand \EOS [0]{\spacefactor3000\relax}%
\providecommand \BibitemShut  [1]{\csname bibitem#1\endcsname}%
\let\auto@bib@innerbib\@empty
\bibitem [{\citenamefont {Barnes}\ \emph {et~al.}(2022)\citenamefont {Barnes}, \citenamefont {Battaglino}, \citenamefont {Bloom}, \citenamefont {Cassella}, \citenamefont {Coxe}, \citenamefont {Crisosto}, \citenamefont {King}, \citenamefont {Kondov}, \citenamefont {Kotru}, \citenamefont {Larsen} \emph {et~al.}}]{barnes2022assembly}%
  \BibitemOpen
  \bibfield  {author} {\bibinfo {author} {\bibfnamefont {K.}~\bibnamefont {Barnes}}, \bibinfo {author} {\bibfnamefont {P.}~\bibnamefont {Battaglino}}, \bibinfo {author} {\bibfnamefont {B.~J.}\ \bibnamefont {Bloom}}, \bibinfo {author} {\bibfnamefont {K.}~\bibnamefont {Cassella}}, \bibinfo {author} {\bibfnamefont {R.}~\bibnamefont {Coxe}}, \bibinfo {author} {\bibfnamefont {N.}~\bibnamefont {Crisosto}}, \bibinfo {author} {\bibfnamefont {J.~P.}\ \bibnamefont {King}}, \bibinfo {author} {\bibfnamefont {S.~S.}\ \bibnamefont {Kondov}}, \bibinfo {author} {\bibfnamefont {K.}~\bibnamefont {Kotru}}, \bibinfo {author} {\bibfnamefont {S.~C.}\ \bibnamefont {Larsen}}, \emph {et~al.},\ }\bibfield  {title} {\bibinfo {title} {Assembly and coherent control of a register of nuclear spin qubits},\ }\href@noop {} {\bibfield  {journal} {\bibinfo  {journal} {Nature Communications}\ }\textbf {\bibinfo {volume} {13}},\ \bibinfo {pages} {2779} (\bibinfo {year} {2022})}\BibitemShut {NoStop}%
\bibitem [{\citenamefont {Levine}\ \emph {et~al.}(2019)\citenamefont {Levine}, \citenamefont {Keesling}, \citenamefont {Semeghini}, \citenamefont {Omran}, \citenamefont {Wang}, \citenamefont {Ebadi}, \citenamefont {Bernien}, \citenamefont {Greiner}, \citenamefont {Vuleti{\'c}}, \citenamefont {Pichler} \emph {et~al.}}]{levine2019parallel}%
  \BibitemOpen
  \bibfield  {author} {\bibinfo {author} {\bibfnamefont {H.}~\bibnamefont {Levine}}, \bibinfo {author} {\bibfnamefont {A.}~\bibnamefont {Keesling}}, \bibinfo {author} {\bibfnamefont {G.}~\bibnamefont {Semeghini}}, \bibinfo {author} {\bibfnamefont {A.}~\bibnamefont {Omran}}, \bibinfo {author} {\bibfnamefont {T.~T.}\ \bibnamefont {Wang}}, \bibinfo {author} {\bibfnamefont {S.}~\bibnamefont {Ebadi}}, \bibinfo {author} {\bibfnamefont {H.}~\bibnamefont {Bernien}}, \bibinfo {author} {\bibfnamefont {M.}~\bibnamefont {Greiner}}, \bibinfo {author} {\bibfnamefont {V.}~\bibnamefont {Vuleti{\'c}}}, \bibinfo {author} {\bibfnamefont {H.}~\bibnamefont {Pichler}}, \emph {et~al.},\ }\bibfield  {title} {\bibinfo {title} {Parallel implementation of high-fidelity multiqubit gates with neutral atoms},\ }\href@noop {} {\bibfield  {journal} {\bibinfo  {journal} {Physical review letters}\ }\textbf {\bibinfo {volume} {123}},\ \bibinfo {pages} {170503} (\bibinfo {year} {2019})}\BibitemShut {NoStop}%
\bibitem [{\citenamefont {Bluvstein}\ \emph {et~al.}(2022)\citenamefont {Bluvstein}, \citenamefont {Levine}, \citenamefont {Semeghini}, \citenamefont {Wang}, \citenamefont {Ebadi}, \citenamefont {Kalinowski}, \citenamefont {Keesling}, \citenamefont {Maskara}, \citenamefont {Pichler}, \citenamefont {Greiner} \emph {et~al.}}]{bluvstein2022quantum}%
  \BibitemOpen
  \bibfield  {author} {\bibinfo {author} {\bibfnamefont {D.}~\bibnamefont {Bluvstein}}, \bibinfo {author} {\bibfnamefont {H.}~\bibnamefont {Levine}}, \bibinfo {author} {\bibfnamefont {G.}~\bibnamefont {Semeghini}}, \bibinfo {author} {\bibfnamefont {T.~T.}\ \bibnamefont {Wang}}, \bibinfo {author} {\bibfnamefont {S.}~\bibnamefont {Ebadi}}, \bibinfo {author} {\bibfnamefont {M.}~\bibnamefont {Kalinowski}}, \bibinfo {author} {\bibfnamefont {A.}~\bibnamefont {Keesling}}, \bibinfo {author} {\bibfnamefont {N.}~\bibnamefont {Maskara}}, \bibinfo {author} {\bibfnamefont {H.}~\bibnamefont {Pichler}}, \bibinfo {author} {\bibfnamefont {M.}~\bibnamefont {Greiner}}, \emph {et~al.},\ }\bibfield  {title} {\bibinfo {title} {A quantum processor based on coherent transport of entangled atom arrays},\ }\href@noop {} {\bibfield  {journal} {\bibinfo  {journal} {Nature}\ }\textbf {\bibinfo {volume} {604}},\ \bibinfo {pages} {451} (\bibinfo {year} {2022})}\BibitemShut {NoStop}%
\bibitem [{\citenamefont {Graham}\ \emph {et~al.}(2022)\citenamefont {Graham}, \citenamefont {Song}, \citenamefont {Scott}, \citenamefont {Poole}, \citenamefont {Phuttitarn}, \citenamefont {Jooya}, \citenamefont {Eichler}, \citenamefont {Jiang}, \citenamefont {Marra}, \citenamefont {Grinkemeyer} \emph {et~al.}}]{graham2022multi}%
  \BibitemOpen
  \bibfield  {author} {\bibinfo {author} {\bibfnamefont {T.}~\bibnamefont {Graham}}, \bibinfo {author} {\bibfnamefont {Y.}~\bibnamefont {Song}}, \bibinfo {author} {\bibfnamefont {J.}~\bibnamefont {Scott}}, \bibinfo {author} {\bibfnamefont {C.}~\bibnamefont {Poole}}, \bibinfo {author} {\bibfnamefont {L.}~\bibnamefont {Phuttitarn}}, \bibinfo {author} {\bibfnamefont {K.}~\bibnamefont {Jooya}}, \bibinfo {author} {\bibfnamefont {P.}~\bibnamefont {Eichler}}, \bibinfo {author} {\bibfnamefont {X.}~\bibnamefont {Jiang}}, \bibinfo {author} {\bibfnamefont {A.}~\bibnamefont {Marra}}, \bibinfo {author} {\bibfnamefont {B.}~\bibnamefont {Grinkemeyer}}, \emph {et~al.},\ }\bibfield  {title} {\bibinfo {title} {Multi-qubit entanglement and algorithms on a neutral-atom quantum computer},\ }\href@noop {} {\bibfield  {journal} {\bibinfo  {journal} {Nature}\ }\textbf {\bibinfo {volume} {604}},\ \bibinfo {pages} {457} (\bibinfo {year} {2022})}\BibitemShut {NoStop}%
\bibitem [{\citenamefont {Ma}\ \emph {et~al.}(2023)\citenamefont {Ma}, \citenamefont {Liu}, \citenamefont {Peng}, \citenamefont {Zhang}, \citenamefont {Jandura}, \citenamefont {Claes}, \citenamefont {Burgers}, \citenamefont {Pupillo}, \citenamefont {Puri},\ and\ \citenamefont {Thompson}}]{ma2023highfidelity}%
  \BibitemOpen
  \bibfield  {author} {\bibinfo {author} {\bibfnamefont {S.}~\bibnamefont {Ma}}, \bibinfo {author} {\bibfnamefont {G.}~\bibnamefont {Liu}}, \bibinfo {author} {\bibfnamefont {P.}~\bibnamefont {Peng}}, \bibinfo {author} {\bibfnamefont {B.}~\bibnamefont {Zhang}}, \bibinfo {author} {\bibfnamefont {S.}~\bibnamefont {Jandura}}, \bibinfo {author} {\bibfnamefont {J.}~\bibnamefont {Claes}}, \bibinfo {author} {\bibfnamefont {A.~P.}\ \bibnamefont {Burgers}}, \bibinfo {author} {\bibfnamefont {G.}~\bibnamefont {Pupillo}}, \bibinfo {author} {\bibfnamefont {S.}~\bibnamefont {Puri}},\ and\ \bibinfo {author} {\bibfnamefont {J.~D.}\ \bibnamefont {Thompson}},\ }\href@noop {} {\bibinfo {title} {High-fidelity gates with mid-circuit erasure conversion in a metastable neutral atom qubit}} (\bibinfo {year} {2023}),\ \Eprint {https://arxiv.org/abs/2305.05493} {arXiv:2305.05493 [quant-ph]} \BibitemShut {NoStop}%
\bibitem [{\citenamefont {Norcia}\ \emph {et~al.}(2023)\citenamefont {Norcia}, \citenamefont {Cairncross}, \citenamefont {Barnes}, \citenamefont {Battaglino}, \citenamefont {Brown}, \citenamefont {Brown}, \citenamefont {Cassella}, \citenamefont {Chen}, \citenamefont {Coxe}, \citenamefont {Crow} \emph {et~al.}}]{norcia2023mid}%
  \BibitemOpen
  \bibfield  {author} {\bibinfo {author} {\bibfnamefont {M.}~\bibnamefont {Norcia}}, \bibinfo {author} {\bibfnamefont {W.}~\bibnamefont {Cairncross}}, \bibinfo {author} {\bibfnamefont {K.}~\bibnamefont {Barnes}}, \bibinfo {author} {\bibfnamefont {P.}~\bibnamefont {Battaglino}}, \bibinfo {author} {\bibfnamefont {A.}~\bibnamefont {Brown}}, \bibinfo {author} {\bibfnamefont {M.}~\bibnamefont {Brown}}, \bibinfo {author} {\bibfnamefont {K.}~\bibnamefont {Cassella}}, \bibinfo {author} {\bibfnamefont {C.-A.}\ \bibnamefont {Chen}}, \bibinfo {author} {\bibfnamefont {R.}~\bibnamefont {Coxe}}, \bibinfo {author} {\bibfnamefont {D.}~\bibnamefont {Crow}}, \emph {et~al.},\ }\bibfield  {title} {\bibinfo {title} {Mid-circuit qubit measurement and rearrangement in a 171 yb atomic array},\ }\href@noop {} {\bibfield  {journal} {\bibinfo  {journal} {arXiv preprint arXiv:2305.19119}\ } (\bibinfo {year} {2023})}\BibitemShut {NoStop}%
\bibitem [{\citenamefont {Graham}\ \emph {et~al.}(2023)\citenamefont {Graham}, \citenamefont {Phuttitarn}, \citenamefont {Chinnarasu}, \citenamefont {Song}, \citenamefont {Poole}, \citenamefont {Jooya}, \citenamefont {Scott}, \citenamefont {Scott}, \citenamefont {Eichler},\ and\ \citenamefont {Saffman}}]{graham2023mid}%
  \BibitemOpen
  \bibfield  {author} {\bibinfo {author} {\bibfnamefont {T.}~\bibnamefont {Graham}}, \bibinfo {author} {\bibfnamefont {L.}~\bibnamefont {Phuttitarn}}, \bibinfo {author} {\bibfnamefont {R.}~\bibnamefont {Chinnarasu}}, \bibinfo {author} {\bibfnamefont {Y.}~\bibnamefont {Song}}, \bibinfo {author} {\bibfnamefont {C.}~\bibnamefont {Poole}}, \bibinfo {author} {\bibfnamefont {K.}~\bibnamefont {Jooya}}, \bibinfo {author} {\bibfnamefont {J.}~\bibnamefont {Scott}}, \bibinfo {author} {\bibfnamefont {A.}~\bibnamefont {Scott}}, \bibinfo {author} {\bibfnamefont {P.}~\bibnamefont {Eichler}},\ and\ \bibinfo {author} {\bibfnamefont {M.}~\bibnamefont {Saffman}},\ }\bibfield  {title} {\bibinfo {title} {Mid-circuit measurements on a neutral atom quantum processor},\ }\href@noop {} {\bibfield  {journal} {\bibinfo  {journal} {arXiv preprint arXiv:2303.10051}\ } (\bibinfo {year} {2023})}\BibitemShut {NoStop}%
\bibitem [{\citenamefont {Lis}\ \emph {et~al.}(2023)\citenamefont {Lis}, \citenamefont {Senoo}, \citenamefont {McGrew}, \citenamefont {R{\"o}nchen}, \citenamefont {Jenkins},\ and\ \citenamefont {Kaufman}}]{lis2023mid}%
  \BibitemOpen
  \bibfield  {author} {\bibinfo {author} {\bibfnamefont {J.~W.}\ \bibnamefont {Lis}}, \bibinfo {author} {\bibfnamefont {A.}~\bibnamefont {Senoo}}, \bibinfo {author} {\bibfnamefont {W.~F.}\ \bibnamefont {McGrew}}, \bibinfo {author} {\bibfnamefont {F.}~\bibnamefont {R{\"o}nchen}}, \bibinfo {author} {\bibfnamefont {A.}~\bibnamefont {Jenkins}},\ and\ \bibinfo {author} {\bibfnamefont {A.~M.}\ \bibnamefont {Kaufman}},\ }\bibfield  {title} {\bibinfo {title} {Mid-circuit operations using the omg-architecture in neutral atom arrays},\ }\href@noop {} {\bibfield  {journal} {\bibinfo  {journal} {arXiv preprint arXiv:2305.19266}\ } (\bibinfo {year} {2023})}\BibitemShut {NoStop}%
\bibitem [{\citenamefont {Deist}\ \emph {et~al.}(2022)\citenamefont {Deist}, \citenamefont {Lu}, \citenamefont {Ho}, \citenamefont {Pasha}, \citenamefont {Zeiher}, \citenamefont {Yan},\ and\ \citenamefont {Stamper-Kurn}}]{PhysRevLett.129.203602}%
  \BibitemOpen
  \bibfield  {author} {\bibinfo {author} {\bibfnamefont {E.}~\bibnamefont {Deist}}, \bibinfo {author} {\bibfnamefont {Y.-H.}\ \bibnamefont {Lu}}, \bibinfo {author} {\bibfnamefont {J.}~\bibnamefont {Ho}}, \bibinfo {author} {\bibfnamefont {M.~K.}\ \bibnamefont {Pasha}}, \bibinfo {author} {\bibfnamefont {J.}~\bibnamefont {Zeiher}}, \bibinfo {author} {\bibfnamefont {Z.}~\bibnamefont {Yan}},\ and\ \bibinfo {author} {\bibfnamefont {D.~M.}\ \bibnamefont {Stamper-Kurn}},\ }\bibfield  {title} {\bibinfo {title} {Mid-circuit cavity measurement in a neutral atom array},\ }\href {https://doi.org/10.1103/PhysRevLett.129.203602} {\bibfield  {journal} {\bibinfo  {journal} {Phys. Rev. Lett.}\ }\textbf {\bibinfo {volume} {129}},\ \bibinfo {pages} {203602} (\bibinfo {year} {2022})}\BibitemShut {NoStop}%
\bibitem [{\citenamefont {Bharti}\ \emph {et~al.}(2022)\citenamefont {Bharti}, \citenamefont {Cervera-Lierta}, \citenamefont {Kyaw}, \citenamefont {Haug}, \citenamefont {Alperin-Lea}, \citenamefont {Anand}, \citenamefont {Degroote}, \citenamefont {Heimonen}, \citenamefont {Kottmann}, \citenamefont {Menke} \emph {et~al.}}]{bharti2022noisy}%
  \BibitemOpen
  \bibfield  {author} {\bibinfo {author} {\bibfnamefont {K.}~\bibnamefont {Bharti}}, \bibinfo {author} {\bibfnamefont {A.}~\bibnamefont {Cervera-Lierta}}, \bibinfo {author} {\bibfnamefont {T.~H.}\ \bibnamefont {Kyaw}}, \bibinfo {author} {\bibfnamefont {T.}~\bibnamefont {Haug}}, \bibinfo {author} {\bibfnamefont {S.}~\bibnamefont {Alperin-Lea}}, \bibinfo {author} {\bibfnamefont {A.}~\bibnamefont {Anand}}, \bibinfo {author} {\bibfnamefont {M.}~\bibnamefont {Degroote}}, \bibinfo {author} {\bibfnamefont {H.}~\bibnamefont {Heimonen}}, \bibinfo {author} {\bibfnamefont {J.~S.}\ \bibnamefont {Kottmann}}, \bibinfo {author} {\bibfnamefont {T.}~\bibnamefont {Menke}}, \emph {et~al.},\ }\bibfield  {title} {\bibinfo {title} {Noisy intermediate-scale quantum algorithms},\ }\href@noop {} {\bibfield  {journal} {\bibinfo  {journal} {Reviews of Modern Physics}\ }\textbf {\bibinfo {volume} {94}},\ \bibinfo {pages} {015004} (\bibinfo {year} {2022})}\BibitemShut {NoStop}%
\bibitem [{\citenamefont {{\DH}or{\dj}evi{\'c}}\ \emph {et~al.}(2021)\citenamefont {{\DH}or{\dj}evi{\'c}}, \citenamefont {Samutpraphoot}, \citenamefont {Ocola}, \citenamefont {Bernien}, \citenamefont {Grinkemeyer}, \citenamefont {Dimitrova}, \citenamefont {Vuleti{\'c}},\ and\ \citenamefont {Lukin}}]{dhordjevic2021entanglement}%
  \BibitemOpen
  \bibfield  {author} {\bibinfo {author} {\bibfnamefont {T.}~\bibnamefont {{\DH}or{\dj}evi{\'c}}}, \bibinfo {author} {\bibfnamefont {P.}~\bibnamefont {Samutpraphoot}}, \bibinfo {author} {\bibfnamefont {P.~L.}\ \bibnamefont {Ocola}}, \bibinfo {author} {\bibfnamefont {H.}~\bibnamefont {Bernien}}, \bibinfo {author} {\bibfnamefont {B.}~\bibnamefont {Grinkemeyer}}, \bibinfo {author} {\bibfnamefont {I.}~\bibnamefont {Dimitrova}}, \bibinfo {author} {\bibfnamefont {V.}~\bibnamefont {Vuleti{\'c}}},\ and\ \bibinfo {author} {\bibfnamefont {M.~D.}\ \bibnamefont {Lukin}},\ }\bibfield  {title} {\bibinfo {title} {Entanglement transport and a nanophotonic interface for atoms in optical tweezers},\ }\href@noop {} {\bibfield  {journal} {\bibinfo  {journal} {Science}\ }\textbf {\bibinfo {volume} {373}},\ \bibinfo {pages} {1511} (\bibinfo {year} {2021})}\BibitemShut {NoStop}%
\bibitem [{\citenamefont {Polloreno}\ and\ \citenamefont {Smith}(2023)}]{polloreno2023qaoa}%
  \BibitemOpen
  \bibfield  {author} {\bibinfo {author} {\bibfnamefont {A.~M.}\ \bibnamefont {Polloreno}}\ and\ \bibinfo {author} {\bibfnamefont {G.}~\bibnamefont {Smith}},\ }\href@noop {} {\bibinfo {title} {The qaoa with few measurements}} (\bibinfo {year} {2023}),\ \Eprint {https://arxiv.org/abs/2205.06845} {arXiv:2205.06845 [quant-ph]} \BibitemShut {NoStop}%
\bibitem [{\citenamefont {Foss-Feig}\ \emph {et~al.}(2023)\citenamefont {Foss-Feig}, \citenamefont {Tikku}, \citenamefont {Lu}, \citenamefont {Mayer}, \citenamefont {Iqbal}, \citenamefont {Gatterman}, \citenamefont {Gerber}, \citenamefont {Gilmore}, \citenamefont {Gresh}, \citenamefont {Hankin} \emph {et~al.}}]{foss2023experimental}%
  \BibitemOpen
  \bibfield  {author} {\bibinfo {author} {\bibfnamefont {M.}~\bibnamefont {Foss-Feig}}, \bibinfo {author} {\bibfnamefont {A.}~\bibnamefont {Tikku}}, \bibinfo {author} {\bibfnamefont {T.-C.}\ \bibnamefont {Lu}}, \bibinfo {author} {\bibfnamefont {K.}~\bibnamefont {Mayer}}, \bibinfo {author} {\bibfnamefont {M.}~\bibnamefont {Iqbal}}, \bibinfo {author} {\bibfnamefont {T.~M.}\ \bibnamefont {Gatterman}}, \bibinfo {author} {\bibfnamefont {J.~A.}\ \bibnamefont {Gerber}}, \bibinfo {author} {\bibfnamefont {K.}~\bibnamefont {Gilmore}}, \bibinfo {author} {\bibfnamefont {D.}~\bibnamefont {Gresh}}, \bibinfo {author} {\bibfnamefont {A.}~\bibnamefont {Hankin}}, \emph {et~al.},\ }\bibfield  {title} {\bibinfo {title} {Experimental demonstration of the advantage of adaptive quantum circuits},\ }\href@noop {} {\bibfield  {journal} {\bibinfo  {journal} {arXiv preprint arXiv:2302.03029}\ } (\bibinfo {year} {2023})}\BibitemShut {NoStop}%
\bibitem [{\citenamefont {Bravyi}\ \emph {et~al.}(2022)\citenamefont {Bravyi}, \citenamefont {Kim}, \citenamefont {Kliesch},\ and\ \citenamefont {Koenig}}]{bravyi2022adaptive}%
  \BibitemOpen
  \bibfield  {author} {\bibinfo {author} {\bibfnamefont {S.}~\bibnamefont {Bravyi}}, \bibinfo {author} {\bibfnamefont {I.}~\bibnamefont {Kim}}, \bibinfo {author} {\bibfnamefont {A.}~\bibnamefont {Kliesch}},\ and\ \bibinfo {author} {\bibfnamefont {R.}~\bibnamefont {Koenig}},\ }\bibfield  {title} {\bibinfo {title} {Adaptive constant-depth circuits for manipulating non-abelian anyons},\ }\href@noop {} {\bibfield  {journal} {\bibinfo  {journal} {arXiv preprint arXiv:2205.01933}\ } (\bibinfo {year} {2022})}\BibitemShut {NoStop}%
\bibitem [{\citenamefont {Yu}\ \emph {et~al.}(2021)\citenamefont {Yu}, \citenamefont {Wilczek},\ and\ \citenamefont {Wu}}]{yu2021quantum}%
  \BibitemOpen
  \bibfield  {author} {\bibinfo {author} {\bibfnamefont {H.}~\bibnamefont {Yu}}, \bibinfo {author} {\bibfnamefont {F.}~\bibnamefont {Wilczek}},\ and\ \bibinfo {author} {\bibfnamefont {B.}~\bibnamefont {Wu}},\ }\bibfield  {title} {\bibinfo {title} {Quantum algorithm for approximating maximum independent sets},\ }\href@noop {} {\bibfield  {journal} {\bibinfo  {journal} {Chinese Physics Letters}\ }\textbf {\bibinfo {volume} {38}},\ \bibinfo {pages} {030304} (\bibinfo {year} {2021})}\BibitemShut {NoStop}%
\bibitem [{\citenamefont {Lucas}(2014)}]{lucas2014ising}%
  \BibitemOpen
  \bibfield  {author} {\bibinfo {author} {\bibfnamefont {A.}~\bibnamefont {Lucas}},\ }\bibfield  {title} {\bibinfo {title} {Ising formulations of many np problems},\ }\href@noop {} {\bibfield  {journal} {\bibinfo  {journal} {Frontiers in physics}\ }\textbf {\bibinfo {volume} {2}},\ \bibinfo {pages} {5} (\bibinfo {year} {2014})}\BibitemShut {NoStop}%
\bibitem [{\citenamefont {Gottesman}(1997)}]{gottesman1997stabilizer}%
  \BibitemOpen
  \bibfield  {author} {\bibinfo {author} {\bibfnamefont {D.}~\bibnamefont {Gottesman}},\ }\href@noop {} {\emph {\bibinfo {title} {Stabilizer codes and quantum error correction}}}\ (\bibinfo  {publisher} {California Institute of Technology},\ \bibinfo {year} {1997})\BibitemShut {NoStop}%
\bibitem [{\citenamefont {Gottesman}(1996)}]{gottesman1996class}%
  \BibitemOpen
  \bibfield  {author} {\bibinfo {author} {\bibfnamefont {D.}~\bibnamefont {Gottesman}},\ }\bibfield  {title} {\bibinfo {title} {Class of quantum error-correcting codes saturating the quantum hamming bound},\ }\href@noop {} {\bibfield  {journal} {\bibinfo  {journal} {Physical Review A}\ }\textbf {\bibinfo {volume} {54}},\ \bibinfo {pages} {1862} (\bibinfo {year} {1996})}\BibitemShut {NoStop}%
\bibitem [{\citenamefont {Li}\ \emph {et~al.}(2023)\citenamefont {Li}, \citenamefont {Lin}, \citenamefont {Zhao}, \citenamefont {Chen},\ and\ \citenamefont {Xia}}]{li2023one}%
  \BibitemOpen
  \bibfield  {author} {\bibinfo {author} {\bibfnamefont {Y.}~\bibnamefont {Li}}, \bibinfo {author} {\bibfnamefont {Z.-P.}\ \bibnamefont {Lin}}, \bibinfo {author} {\bibfnamefont {X.-Y.}\ \bibnamefont {Zhao}}, \bibinfo {author} {\bibfnamefont {Y.-H.}\ \bibnamefont {Chen}},\ and\ \bibinfo {author} {\bibfnamefont {Y.}~\bibnamefont {Xia}},\ }\bibfield  {title} {\bibinfo {title} {One-step implementation of multiqubit controlled--controlled-z gates with rydberg atoms},\ }\href@noop {} {\bibfield  {journal} {\bibinfo  {journal} {Laser Physics Letters}\ }\textbf {\bibinfo {volume} {20}},\ \bibinfo {pages} {095205} (\bibinfo {year} {2023})}\BibitemShut {NoStop}%
\bibitem [{\citenamefont {Henriet}\ \emph {et~al.}(2020)\citenamefont {Henriet}, \citenamefont {Beguin}, \citenamefont {Signoles}, \citenamefont {Lahaye}, \citenamefont {Browaeys}, \citenamefont {Reymond},\ and\ \citenamefont {Jurczak}}]{henriet2020quantum}%
  \BibitemOpen
  \bibfield  {author} {\bibinfo {author} {\bibfnamefont {L.}~\bibnamefont {Henriet}}, \bibinfo {author} {\bibfnamefont {L.}~\bibnamefont {Beguin}}, \bibinfo {author} {\bibfnamefont {A.}~\bibnamefont {Signoles}}, \bibinfo {author} {\bibfnamefont {T.}~\bibnamefont {Lahaye}}, \bibinfo {author} {\bibfnamefont {A.}~\bibnamefont {Browaeys}}, \bibinfo {author} {\bibfnamefont {G.-O.}\ \bibnamefont {Reymond}},\ and\ \bibinfo {author} {\bibfnamefont {C.}~\bibnamefont {Jurczak}},\ }\bibfield  {title} {\bibinfo {title} {Quantum computing with neutral atoms},\ }\href@noop {} {\bibfield  {journal} {\bibinfo  {journal} {Quantum}\ }\textbf {\bibinfo {volume} {4}},\ \bibinfo {pages} {327} (\bibinfo {year} {2020})}\BibitemShut {NoStop}%
\bibitem [{\citenamefont {M{\"u}ller}\ \emph {et~al.}(2009)\citenamefont {M{\"u}ller}, \citenamefont {Lesanovsky}, \citenamefont {Weimer}, \citenamefont {B{\"u}chler},\ and\ \citenamefont {Zoller}}]{muller2009mesoscopic}%
  \BibitemOpen
  \bibfield  {author} {\bibinfo {author} {\bibfnamefont {M.}~\bibnamefont {M{\"u}ller}}, \bibinfo {author} {\bibfnamefont {I.}~\bibnamefont {Lesanovsky}}, \bibinfo {author} {\bibfnamefont {H.}~\bibnamefont {Weimer}}, \bibinfo {author} {\bibfnamefont {H.}~\bibnamefont {B{\"u}chler}},\ and\ \bibinfo {author} {\bibfnamefont {P.}~\bibnamefont {Zoller}},\ }\bibfield  {title} {\bibinfo {title} {Mesoscopic rydberg gate based on electromagnetically induced transparency},\ }\href@noop {} {\bibfield  {journal} {\bibinfo  {journal} {Physical Review Letters}\ }\textbf {\bibinfo {volume} {102}},\ \bibinfo {pages} {170502} (\bibinfo {year} {2009})}\BibitemShut {NoStop}%
\bibitem [{\citenamefont {Guo}\ \emph {et~al.}(2019)\citenamefont {Guo}, \citenamefont {Jerrum},\ and\ \citenamefont {Liu}}]{guo2019uniform}%
  \BibitemOpen
  \bibfield  {author} {\bibinfo {author} {\bibfnamefont {H.}~\bibnamefont {Guo}}, \bibinfo {author} {\bibfnamefont {M.}~\bibnamefont {Jerrum}},\ and\ \bibinfo {author} {\bibfnamefont {J.}~\bibnamefont {Liu}},\ }\bibfield  {title} {\bibinfo {title} {Uniform sampling through the lov{\'a}sz local lemma},\ }\href@noop {} {\bibfield  {journal} {\bibinfo  {journal} {Journal of the ACM (JACM)}\ }\textbf {\bibinfo {volume} {66}},\ \bibinfo {pages} {1} (\bibinfo {year} {2019})}\BibitemShut {NoStop}%
\bibitem [{\citenamefont {Jerrum}(2021)}]{jerrum2021fundamentals}%
  \BibitemOpen
  \bibfield  {author} {\bibinfo {author} {\bibfnamefont {M.}~\bibnamefont {Jerrum}},\ }\bibfield  {title} {\bibinfo {title} {Fundamentals of partial rejection sampling},\ }\href@noop {} {\bibfield  {journal} {\bibinfo  {journal} {arXiv preprint arXiv:2106.07744}\ } (\bibinfo {year} {2021})}\BibitemShut {NoStop}%
\bibitem [{\citenamefont {Sly}\ and\ \citenamefont {Sun}(2012)}]{sly2012computational}%
  \BibitemOpen
  \bibfield  {author} {\bibinfo {author} {\bibfnamefont {A.}~\bibnamefont {Sly}}\ and\ \bibinfo {author} {\bibfnamefont {N.}~\bibnamefont {Sun}},\ }\bibfield  {title} {\bibinfo {title} {The computational hardness of counting in two-spin models on d-regular graphs},\ }in\ \href@noop {} {\emph {\bibinfo {booktitle} {2012 IEEE 53rd Annual Symposium on Foundations of Computer Science}}}\ (\bibinfo {organization} {IEEE},\ \bibinfo {year} {2012})\ pp.\ \bibinfo {pages} {361--369}\BibitemShut {NoStop}%
\bibitem [{\citenamefont {Weitz}(2006)}]{weitz2006counting}%
  \BibitemOpen
  \bibfield  {author} {\bibinfo {author} {\bibfnamefont {D.}~\bibnamefont {Weitz}},\ }\bibfield  {title} {\bibinfo {title} {Counting independent sets up to the tree threshold},\ }in\ \href@noop {} {\emph {\bibinfo {booktitle} {Proceedings of the thirty-eighth annual ACM symposium on Theory of computing}}}\ (\bibinfo {year} {2006})\ pp.\ \bibinfo {pages} {140--149}\BibitemShut {NoStop}%
\bibitem [{\citenamefont {Galanis}\ \emph {et~al.}(2016)\citenamefont {Galanis}, \citenamefont {{\v{S}}tefankovi{\v{c}}},\ and\ \citenamefont {Vigoda}}]{galanis2016inapproximability}%
  \BibitemOpen
  \bibfield  {author} {\bibinfo {author} {\bibfnamefont {A.}~\bibnamefont {Galanis}}, \bibinfo {author} {\bibfnamefont {D.}~\bibnamefont {{\v{S}}tefankovi{\v{c}}}},\ and\ \bibinfo {author} {\bibfnamefont {E.}~\bibnamefont {Vigoda}},\ }\bibfield  {title} {\bibinfo {title} {Inapproximability of the partition function for the antiferromagnetic ising and hard-core models},\ }\href@noop {} {\bibfield  {journal} {\bibinfo  {journal} {Combinatorics, Probability and Computing}\ }\textbf {\bibinfo {volume} {25}},\ \bibinfo {pages} {500} (\bibinfo {year} {2016})}\BibitemShut {NoStop}%
\bibitem [{\citenamefont {Grover}(1996)}]{grover1996fast}%
  \BibitemOpen
  \bibfield  {author} {\bibinfo {author} {\bibfnamefont {L.~K.}\ \bibnamefont {Grover}},\ }\bibfield  {title} {\bibinfo {title} {A fast quantum mechanical algorithm for database search},\ }in\ \href@noop {} {\emph {\bibinfo {booktitle} {Proceedings of the twenty-eighth annual ACM symposium on Theory of computing}}}\ (\bibinfo {year} {1996})\ pp.\ \bibinfo {pages} {212--219}\BibitemShut {NoStop}%
\bibitem [{\citenamefont {Brassard}\ \emph {et~al.}(1998)\citenamefont {Brassard}, \citenamefont {H{\o}yer},\ and\ \citenamefont {Tapp}}]{brassard1998quantum}%
  \BibitemOpen
  \bibfield  {author} {\bibinfo {author} {\bibfnamefont {G.}~\bibnamefont {Brassard}}, \bibinfo {author} {\bibfnamefont {P.}~\bibnamefont {H{\o}yer}},\ and\ \bibinfo {author} {\bibfnamefont {A.}~\bibnamefont {Tapp}},\ }\bibfield  {title} {\bibinfo {title} {Quantum counting},\ }in\ \href@noop {} {\emph {\bibinfo {booktitle} {Automata, Languages and Programming\: 25th International Colloquium, ICALP\'98 Aalborg, Denmark, July 13--17, 1998 Proceedings 25}}}\ (\bibinfo {organization} {Springer},\ \bibinfo {year} {1998})\ pp.\ \bibinfo {pages} {820--831}\BibitemShut {NoStop}%
\bibitem [{\citenamefont {Bravyi}\ \emph {et~al.}(2011)\citenamefont {Bravyi}, \citenamefont {Harrow},\ and\ \citenamefont {Hassidim}}]{bravyi2011quantum}%
  \BibitemOpen
  \bibfield  {author} {\bibinfo {author} {\bibfnamefont {S.}~\bibnamefont {Bravyi}}, \bibinfo {author} {\bibfnamefont {A.~W.}\ \bibnamefont {Harrow}},\ and\ \bibinfo {author} {\bibfnamefont {A.}~\bibnamefont {Hassidim}},\ }\bibfield  {title} {\bibinfo {title} {Quantum algorithms for testing properties of distributions},\ }\href@noop {} {\bibfield  {journal} {\bibinfo  {journal} {IEEE Transactions on Information Theory}\ }\textbf {\bibinfo {volume} {57}},\ \bibinfo {pages} {3971} (\bibinfo {year} {2011})}\BibitemShut {NoStop}%
\bibitem [{\citenamefont {Wu}\ \emph {et~al.}(2020)\citenamefont {Wu}, \citenamefont {Yu},\ and\ \citenamefont {Wilczek}}]{PhysRevA.101.012318}%
  \BibitemOpen
  \bibfield  {author} {\bibinfo {author} {\bibfnamefont {B.}~\bibnamefont {Wu}}, \bibinfo {author} {\bibfnamefont {H.}~\bibnamefont {Yu}},\ and\ \bibinfo {author} {\bibfnamefont {F.}~\bibnamefont {Wilczek}},\ }\bibfield  {title} {\bibinfo {title} {Quantum independent-set problem and non-abelian adiabatic mixing},\ }\href {https://doi.org/10.1103/PhysRevA.101.012318} {\bibfield  {journal} {\bibinfo  {journal} {Phys. Rev. A}\ }\textbf {\bibinfo {volume} {101}},\ \bibinfo {pages} {012318} (\bibinfo {year} {2020})}\BibitemShut {NoStop}%
\bibitem [{\citenamefont {Wild}\ \emph {et~al.}(2021)\citenamefont {Wild}, \citenamefont {Sels}, \citenamefont {Pichler}, \citenamefont {Zanoci},\ and\ \citenamefont {Lukin}}]{wild2021quantum}%
  \BibitemOpen
  \bibfield  {author} {\bibinfo {author} {\bibfnamefont {D.~S.}\ \bibnamefont {Wild}}, \bibinfo {author} {\bibfnamefont {D.}~\bibnamefont {Sels}}, \bibinfo {author} {\bibfnamefont {H.}~\bibnamefont {Pichler}}, \bibinfo {author} {\bibfnamefont {C.}~\bibnamefont {Zanoci}},\ and\ \bibinfo {author} {\bibfnamefont {M.~D.}\ \bibnamefont {Lukin}},\ }\bibfield  {title} {\bibinfo {title} {Quantum sampling algorithms for near-term devices},\ }\href@noop {} {\bibfield  {journal} {\bibinfo  {journal} {Physical Review Letters}\ }\textbf {\bibinfo {volume} {127}},\ \bibinfo {pages} {100504} (\bibinfo {year} {2021})}\BibitemShut {NoStop}%
\bibitem [{\citenamefont {Brassard}\ \emph {et~al.}(2000)\citenamefont {Brassard}, \citenamefont {Hoyer}, \citenamefont {Mosca},\ and\ \citenamefont {Tapp}}]{brassard2000quantum}%
  \BibitemOpen
  \bibfield  {author} {\bibinfo {author} {\bibfnamefont {G.}~\bibnamefont {Brassard}}, \bibinfo {author} {\bibfnamefont {P.}~\bibnamefont {Hoyer}}, \bibinfo {author} {\bibfnamefont {M.}~\bibnamefont {Mosca}},\ and\ \bibinfo {author} {\bibfnamefont {A.}~\bibnamefont {Tapp}},\ }\bibfield  {title} {\bibinfo {title} {Quantum amplitude amplification and estimation},\ }\href@noop {} {\bibfield  {journal} {\bibinfo  {journal} {Quantum Computation and Quantum Information: A Millennium Volume. AMS Contemporary Mathematics Series}\ } (\bibinfo {year} {2000})}\BibitemShut {NoStop}%
\bibitem [{\citenamefont {Herman}\ \emph {et~al.}(2023)\citenamefont {Herman}, \citenamefont {Shaydulin}, \citenamefont {Sun}, \citenamefont {Chakrabarti}, \citenamefont {Hu}, \citenamefont {Minssen}, \citenamefont {Rattew}, \citenamefont {Yalovetzky},\ and\ \citenamefont {Pistoia}}]{herman2023constrained}%
  \BibitemOpen
  \bibfield  {author} {\bibinfo {author} {\bibfnamefont {D.}~\bibnamefont {Herman}}, \bibinfo {author} {\bibfnamefont {R.}~\bibnamefont {Shaydulin}}, \bibinfo {author} {\bibfnamefont {Y.}~\bibnamefont {Sun}}, \bibinfo {author} {\bibfnamefont {S.}~\bibnamefont {Chakrabarti}}, \bibinfo {author} {\bibfnamefont {S.}~\bibnamefont {Hu}}, \bibinfo {author} {\bibfnamefont {P.}~\bibnamefont {Minssen}}, \bibinfo {author} {\bibfnamefont {A.}~\bibnamefont {Rattew}}, \bibinfo {author} {\bibfnamefont {R.}~\bibnamefont {Yalovetzky}},\ and\ \bibinfo {author} {\bibfnamefont {M.}~\bibnamefont {Pistoia}},\ }\bibfield  {title} {\bibinfo {title} {Constrained optimization via quantum zeno dynamics},\ }\href@noop {} {\bibfield  {journal} {\bibinfo  {journal} {Communications Physics}\ }\textbf {\bibinfo {volume} {6}},\ \bibinfo {pages} {219} (\bibinfo {year} {2023})}\BibitemShut {NoStop}%
\bibitem [{\citenamefont {Sieberer}\ \emph {et~al.}(2019)\citenamefont {Sieberer}, \citenamefont {Olsacher}, \citenamefont {Elben}, \citenamefont {Heyl}, \citenamefont {Hauke}, \citenamefont {Haake},\ and\ \citenamefont {Zoller}}]{sieberer2019digital}%
  \BibitemOpen
  \bibfield  {author} {\bibinfo {author} {\bibfnamefont {L.~M.}\ \bibnamefont {Sieberer}}, \bibinfo {author} {\bibfnamefont {T.}~\bibnamefont {Olsacher}}, \bibinfo {author} {\bibfnamefont {A.}~\bibnamefont {Elben}}, \bibinfo {author} {\bibfnamefont {M.}~\bibnamefont {Heyl}}, \bibinfo {author} {\bibfnamefont {P.}~\bibnamefont {Hauke}}, \bibinfo {author} {\bibfnamefont {F.}~\bibnamefont {Haake}},\ and\ \bibinfo {author} {\bibfnamefont {P.}~\bibnamefont {Zoller}},\ }\bibfield  {title} {\bibinfo {title} {Digital quantum simulation, trotter errors, and quantum chaos of the kicked top},\ }\href@noop {} {\bibfield  {journal} {\bibinfo  {journal} {npj Quantum Information}\ }\textbf {\bibinfo {volume} {5}},\ \bibinfo {pages} {78} (\bibinfo {year} {2019})}\BibitemShut {NoStop}%
\bibitem [{\citenamefont {Heyl}\ \emph {et~al.}(2019)\citenamefont {Heyl}, \citenamefont {Hauke},\ and\ \citenamefont {Zoller}}]{heyl2019quantum}%
  \BibitemOpen
  \bibfield  {author} {\bibinfo {author} {\bibfnamefont {M.}~\bibnamefont {Heyl}}, \bibinfo {author} {\bibfnamefont {P.}~\bibnamefont {Hauke}},\ and\ \bibinfo {author} {\bibfnamefont {P.}~\bibnamefont {Zoller}},\ }\bibfield  {title} {\bibinfo {title} {Quantum localization bounds trotter errors in digital quantum simulation},\ }\href@noop {} {\bibfield  {journal} {\bibinfo  {journal} {Science advances}\ }\textbf {\bibinfo {volume} {5}},\ \bibinfo {pages} {eaau8342} (\bibinfo {year} {2019})}\BibitemShut {NoStop}%
\end{thebibliography}%
\appendix

\clearpage
\newpage

\begin{widetext}

\section{Subspace syndrome detection}\label{app:detections}
In this appendix, we provide a recipe to construct stabilizers for common constraint terms in Ising type problems, as those in the work by Lucas \cite{lucas2014ising}. We follow this with an example application to 1-hot-encoding constraints.

Consider a graph $\mathcal G = (V,E)$ , where $V$ is equipped with one or more binary variables $x_v \in \mathcal X$ over each vertex $v \in V$. Furthermore, we form a map from the set of binary variable values $\{x_v\}$ to the set of bitstrings $\mathcal B$ that form a representation of the state.  
Let us define a set of constraints $\mathcal C$, which are a set of Boolean valued functions $c(\mathcal X)$ that take a collection of variables $\{x_v\}$ and output True or False. Any constraint $c \in \mathcal{C}$  forms a partition on the set of bitstrings $\mathcal B$ into a set $\mathcal V_c$ which evaluate to True and its complement $\bar{\mathcal V}_c$ which evaluate to False.

Then for any problem which can be described by the previous definitions, we can trivially form the stabilizers on the constraint space . For each $c$ there exists an operator $\hat S_c$ defined as follows. Let us define the following single qubit projectors:

\begin{equation}
    P^0 = \begin{pmatrix}
        1 & 0 \\
        0 & 0
    \end{pmatrix} \qquad
        P^1 = \begin{pmatrix}
        0 & 0 \\
        0 & 1
    \end{pmatrix}
\end{equation}

Assume that the bitstring representations of state are length-$k$ bitstrings $b \in \mathcal{B}$ where $b_n$ is the value $(0,1)$ of the $n^{\text{th}}$ place. Then $\hat S_c$ is constructed as:

\begin{equation}\label{eq: stab_recipe}
\hat S_c = \sum_{b\in \mathcal V_c} P_0^{b_0} P_1^{b_1} \cdots P_k^{b_k} - \sum_{b\in \bar {\mathcal V}_c}P_0^{b_0} P_1^{b_1} \cdots P_k^{b_k}
\end{equation}
Such an operator is necessarily diagonal in the computational basis that the constraints are defined in. The collection of these operators mutually commute, even if the constraints are not satisfiable, and similarly form a group. Hence, the set $\{\hat S_c | c\in \mathcal C\}$  stabilizes the constraint subspace.

The syndrome extraction operator onto an ancilla qubit can be formed similarly by replacing the phase with a bitflip on an ancilla:

\begin{equation}
    \hat G_c  = \left(\sum_{b\in \mathcal V_c} P_0^{b_0} P_1^{b_1} \cdots P_k^{b_k} \right)1_{a_c} + \left(\sum_{b\in \bar {\mathcal V}_c}P_0^{b_0} P_1^{b_1} \cdots P_k^{b_k}\right)X_{a_c}
\end{equation}
where $a_c$ is the ancilla qubit state space associated with constraint $c$.

\subsection{Example: 1-hot-encoding constraints}

One of the most common forms of constraints for Ising (QUBO) type problems is the ``one-hot-encoding", which constraints the space of $n$ binary variables such that one, and only one, may be in the $1$ state. This takes the form:

\begin{equation}
    H_A = (1-\sum_i^n Z_i)^2
\end{equation}
The above $H_A$ has energy $E_A >0$ whenever the one-hot constraint is violated. These kinds of constraints appear in many graph theory problems, including minimal covers, graph colorings, and even in local enforcement of tree topologies. 

Consider the $3$-variable version of this constraint. Following the recipe of \eqref{eq: stab_recipe}, the stabilizer is given by
\begin{align*}
    \hat S_c=& \;(P^1_0P^0_1P^0_2 + P^0_0P^1_1P^0_2 +P^0_0P^0_1P^1_2) \\
    &- (P^0_0P^0_1P^0_2 + P^1_0P^1_1P^0_2 +P^0_0P^1_1P^1_2 +P^1_0P^1_1P^1_2 +P^1_0P^1_1P^1_2),
    \end{align*}

with the full unitary matrix form of the operation given by:
\begin{equation}
    \hat S_c =\begin{pmatrix}
-1 &  0 &  0 &  0 &  0 &  0 &  0 &  0 \\
 0 &  1 &  0 &  0 &  0 &  0 &  0 &  0 \\
 0 &  0 &  1 &  0 &  0 &  0 &  0 &  0 \\
 0 &  0 &  0 & -1 &  0 &  0 &  0 &  0 \\
 0 &  0 &  0 &  0 &  1 &  0 &  0 &  0 \\
 0 &  0 &  0 &  0 &  0 & -1 &  0 &  0 \\
 0 &  0 &  0 &  0 &  0 &  0 & -1 &  0 \\
 0 &  0 &  0 &  0 &  0 &  0 &  0 & -1 \\
\end{pmatrix}
\end{equation}

The syndrome extraction is then

\begin{equation}
\begin{aligned}
    \hat G_c=& \;(P^1_0P^0_1P^0_2 + P^0_0P^1_1P^0_2 +P^0_0P^0_1P^1_2)I_a \\
    &+ (P^0_0P^0_1P^0_2 + P^1_0P^1_1P^0_2 +P^0_0P^1_1P^1_2 +P^1_0P^1_1P^1_2 +P^1_0P^1_1P^1_2)X_a
    \end{aligned}
\end{equation}

The full form of the syndrome measurement is of the form:

\begin{equation}
\mathcal G_C = \left(
\begin{array}{*{16}c}
1 & 0 & 0 & 0 & 0 & 0 & 0 & 0 & 0 & 0 & 0 & 0 & 0 & 0 & 0 & 0 \\
0 & 1 & 0 & 0 & 0 & 0 & 0 & 0 & 0 & 0 & 0 & 0 & 0 & 0 & 0 & 0 \\
0 & 0 & 1 & 0 & 0 & 0 & 0 & 0 & 0 & 0 & 0 & 0 & 0 & 0 & 0 & 0 \\
0 & 0 & 0 & 1 & 0 & 0 & 0 & 0 & 0 & 0 & 0 & 0 & 0 & 0 & 0 & 0 \\
0 & 0 & 0 & 0 & 1 & 0 & 0 & 0 & 0 & 0 & 0 & 0 & 0 & 0 & 0 & 0 \\
0 & 0 & 0 & 0 & 0 & 1 & 0 & 0 & 0 & 0 & 0 & 0 & 0 & 0 & 0 & 0 \\
0 & 0 & 0 & 0 & 0 & 0 & 0 & 1 & 0 & 0 & 0 & 0 & 0 & 0 & 0 & 0 \\
0 & 0 & 0 & 0 & 0 & 0 & 1 & 0 & 0 & 0 & 0 & 0 & 0 & 0 & 0 & 0 \\
0 & 0 & 0 & 0 & 0 & 0 & 0 & 0 & 1 & 0 & 0 & 0 & 0 & 0 & 0 & 0 \\
0 & 0 & 0 & 0 & 0 & 0 & 0 & 0 & 0 & 1 & 0 & 0 & 0 & 0 & 0 & 0 \\
0 & 0 & 0 & 0 & 0 & 0 & 0 & 0 & 0 & 0 & 0 & 1 & 0 & 0 & 0 & 0 \\
0 & 0 & 0 & 0 & 0 & 0 & 0 & 0 & 0 & 0 & 1 & 0 & 0 & 0 & 0 & 0 \\
0 & 0 & 0 & 0 & 0 & 0 & 0 & 0 & 0 & 0 & 0 & 0 & 0 & 1 & 0 & 0 \\
0 & 0 & 0 & 0 & 0 & 0 & 0 & 0 & 0 & 0 & 0 & 0 & 1 & 0 & 0 & 0 \\
0 & 0 & 0 & 0 & 0 & 0 & 0 & 0 & 0 & 0 & 0 & 0 & 0 & 0 & 0 & 1 \\
0 & 0 & 0 & 0 & 0 & 0 & 0 & 0 & 0 & 0 & 0 & 0 & 0 & 0 & 1 & 0 
\end{array}
\right)
\end{equation}

which can be compiled into a sequence of native gates as desired.

\section{State preparation proof}\label{sec:Proof}

\noindent\textbf{Protocol.} Assign a qubit to each vertex of a graph $\mathcal{G}=(V, E)$. Initialize the system in the state
\begin{align}
	\psi_\text{init}=\bigotimes_{v\in V}\psi_v\hspace{30pt}\psi_v=\sqrt{1-p_v}\ket{0}+\sqrt{p_v}\ket{1}
\end{align}
for some set of vertex probabilities $0<p_v<1$. For each edge $e\in E$, perform the projective measurement
\begin{align}
	M^e_1&=\ketbra{11}{11}_e\otimes \mathbf{1}_{v\notin e}\\
	M^e_0&=1-M_1^e.
\end{align}
Let $B$ be the set of all vertices that are part of any edge for which $M^e=1$ and let $\mathcal{N}(B)$ be the neighborhood of $B$. For each $v\in B\cup \mathcal{N}(B)$, reset the corresponding qubit to the state $\psi_v$. Repeat this measurement and reset process until there are no violating edges.\\
\\
\noindent\textbf{Claim.} If the protocol described above terminates, the system is prepared in the state
\begin{align}
	\psi_\text{targ}\propto\sum_{\mathbf{s}\in \mathcal{I}(A)}\prod_{v\in V}\sqrt{p_v^{s_v}(1-p_v)^{1-s_v}}\ket{\mathbf{s}},
\end{align}
where $\mathcal{I}(V)$ is the set of independent sets of $\mathcal{G}$. Moreover, the distribution of the number of measure-and-reset steps it takes to prepare this state is equal to the distribution over the number of iterations it would take to detect no violations if the initial and reset states are replaced by classical random variables drawn from the corresponding distributions, i.e if the state preparation procedure were converted into a classical sampling procedure.\\
\\
\noindent 
\textbf{Proof.} For some $A\subseteq V$, suppose that the system is in the state
\begin{align}
\psi_A=\sum_{\mathbf{s}\in \mathcal{I}(A)}\alpha_\mathbf{s}\ket{\mathbf{s}}\otimes\bigotimes_{v\in V\setminus A}\psi_v,
\end{align}
where $\mathcal{I}(A)$ is the set of independent sets on the subgraph induced by $A$, and
\begin{align}
\alpha_\mathbf{s}\propto \prod_{v\in A}\sqrt{p_v^{s_v}(1-p_v)^{1-s_v}}.
\end{align}
Consider performing projective measurements 
\begin{align}
	M^e_1&=\ketbra{11}{11}_e\otimes \mathbf{1}_{v\notin e}\\
	M^e_0&=1-M_1^e
\end{align}
for each edge $e$. Let $B$ be the set of all vertices that are part of any edge for which $M^e=1$. Let $\mathcal{N}(B)$ be the neighborhood of $B$. Define $A'=V\setminus(B\cup \mathcal{N}(B))$. After measurement, all qubits in $B$ must be in the state $\ket{1}$, and all qubits in the neighborhood $\mathcal{N}(B)$ must be in the state $\ket{0}$, so that we obtain the state
\begin{align}
\psi_{A'}=\sum_{\mathbf{s}'\in \mathcal{S}(A')}\alpha'_{\mathbf{s}'}\ket{\mathbf{s}'}\otimes\ket{0}_{\mathcal{N}(B)}\otimes\ket{1}_B,
\end{align}
where $\mathcal{S}(A')$ is the set of all vertex assignments of $A'$. For all $\mathbf{s}'\in\mathcal{S}(A')$ such that $\alpha'_{\mathbf{s}'}\neq 0$, $\mathbf{s}'$ must be an independent set of the subgraph induced by $A'$, as otherwise there would be vertices in $A'$ corresponding to edges violating the independence constraint, contradicting the definition of $A'$. Moreover, for any independent set $\mathbf{s}'\in\mathcal{I}(A')$, consider the vertex assignment $\tilde{\mathbf{s}}'$ on $A$ defined by
\begin{align}
\tilde{\mathbf{s}}'_v=\begin{cases}
\mathbf{s}'_v & v\in A'\\
0 & v\in \mathcal{N}(B)\\
1 & v\in B.
\end{cases}
\end{align}
This is an independent set on the subgraph induced by $A$. To see that this is the case, consider two vertices of $A$. If one of them is in $\mathcal{N}(B)$, then it is set to zero and so cannot participate in a violation of independence. Otherwise, suppose that both are in $A'$. Then they do not violate independence by the assumption that $\mathbf{s}'$ is independent in the subgraph induced by $A'$. Both cannot be in $B$, as this would violate the assumption that the initial state restricted to $A$ had support only on independent sets. Finally, if one is in $A$ and the other is in $B$, then they cannot share an edge, because by definition $A\cap \mathcal{N}(B)=\emptyset$. This is the only independent set on $A$ that is in the support of the projectors applied by the measurement with the obtained outcome and restricts to $\mathbf{s}'$ on $A'$. Therefore, we have
\begin{align}
\psi_{A'}=\sum_{\mathbf{s}'\in \mathcal{I}(A')}\prod_{v\in A'\setminus A}\sqrt{p_v^{s'_v}(1-p_v)^{1-s'_v}}\,\alpha_{\tilde{\mathbf{s}}'}\ket{\mathbf{s}'}\otimes\ket{0}_{\mathcal{N}(B)}\otimes\ket{1}_B.
\end{align}
Because 
for any two $A'$-independent sets $\mathbf{s}'$, $\mathbf{r}'$, the restrictions of $\tilde{\mathbf{s}}'$ and $\tilde{\mathbf{r}}'$ to $A\setminus A'$ are the same we have
\begin{align}
\alpha_{\tilde{\mathbf{s}}'}\propto \prod_{v\in A'}\sqrt{p_v^{s'_v}(1-p_v)^{1-s'_v}}.
\end{align}
Thus resetting the qubits in $B$ and $\mathcal{N}(B)$ as prescribed, we obtain the state

\begin{align}
\psi_{A'}=\sum_{\mathbf{s}'\in \mathcal{I}(A')}\prod_{v\in A'}\sqrt{p_v^{s_v}(1-p_v)^{1-s_v}}\,\ket{\mathbf{s}'}\otimes\bigotimes_{v\in V\setminus A'}\psi_v,
\end{align}
Noting that the initial state $\psi_\text{init}=\psi_A$ for $A=\emptyset$ and the target state $\psi_\text{targ}=\psi_A$ for $A=V$, we conclude that the target state is prepared once the measurement $B=\emptyset$ (no violating edges) is obtained.\\
\\
To see that the classical process described above has the same statistics as the state preparation process, note that the latter consists simply of repeated application of a quantum instrument whose elements have the form
\begin{align}
\mathcal{M}_{B}=\mathbf{1}_{v\notin B\cup\mathbf{N}(B)}\otimes \left(\bigotimes_{v\in B\cup\mathcal{N}(B)}\psi_v\right)\left(\bra{0}_{\mathcal{N}(B)}\otimes\bra{1}_B\right).
\end{align} 
Denoting the total decoherence channel in the computational basis by $\mathcal{D}$, we have
\begin{align}
\mathcal{D}\circ\mathcal{M}_B=\mathcal{D}\circ\mathcal{M}_B\circ\mathcal{D}.
\end{align}
The probability of observing the sequence of measurements $B_1, B_2, \ldots, B_m$ and then obtaining the bitstring $\mathbf{s}$ on measuring all qubits in the computational basis is
\begin{align}
\text{Pr}(B_1, B_2,\ldots,B_m,\mathbf{s})&=\bra{\mathbf{s}}\mathcal{M}_{B_m}\circ\mathcal{M}_{B_{m-1}}\circ\ldots\circ\mathcal{M}_{B_1}(\rho_\text{init})\ket{\mathbf{s}}\\
&=\bra{\mathbf{s}}\mathcal{D}\circ\mathcal{M}_{B_m}\circ\mathcal{M}_{B_{m-1}}\circ\ldots\circ\mathcal{M}_{B_1}(\rho_\text{init})\ket{\mathbf{s}}\\
&=\bra{\mathbf{s}}\mathcal{D}\circ\mathcal{M}_{B_m}\circ\mathcal{D}\circ\mathcal{M}_{B_{m-1}}\circ\ldots\circ\mathcal{D}\circ\mathcal{M}_{B_1}\circ\mathcal{D}(\rho_\text{init})\ket{\mathbf{s}}.
\end{align}
The process described by the operator acting on $\rho_\text{init}$ in the last line is identical to the classical process described.

\section{General 2Q Recovery Sequence}\label{sec:sequence}

A general two-qubit recovery sequence to prepare a state $|\psi\rangle = \alpha|00\rangle + \beta|01\rangle + \gamma |10\rangle + \eta|11\rangle$ starting from $|00\rangle$,  where $\alpha, \beta, \gamma, \eta$ are complex amplitudes with unit norm, the following gate sequence suffices:

\begin{center}
\begin{quantikz}
   \lstick{$|0\rangle$} & \gate{U_1} & \ctrl{1}  &\qw  \\
   \lstick{$|0\rangle$} &\gate{U_2}  & \gate{U}  &\qw 
\end{quantikz}
\end{center}

We define the above gates via the following set of relationships:
\begin{equation}
    U_2 = R_z\left(\frac{\theta_\alpha -\theta_\beta}{2}\right) R_y(\pi+\theta)R_z(\frac{\theta_\alpha + \theta_\beta}{2})
\end{equation}
where $\theta = 2\arctan\left(\frac{|\alpha|}{|\beta|}\right)$. We also have

\begin{equation}
    U_1 = R_y(\pi + \theta) 
\end{equation}
where 
\begin{equation}
    \theta = 2 \arctan\left( \sqrt{\frac{|\alpha|^2 + |\beta|^2}{|\gamma|^2 + |\eta|^2}} \right)
\end{equation}

In some compiling contexts, it can be useful can further break down this sequence as the following:

\begin{center}
    \begin{quantikz}
    \lstick{$|0\rangle$} & \gate{U_1} & \ctrl{1} & \qw       &  \ctrl{1} &  \qw  &  \qw   &  \qw  \\
    \lstick{$|0\rangle$} & \gate{A'}  & \targ{}  &  \gate{B} &  \targ{}  &  \qw  &  \gate{C} &  \qw \\
    \end{quantikz}
\end{center}

where $A' = A U_2$ and 
\begin{equation}
    A = R_z\left(\tilde\alpha\right) R_y\left(\frac{\tilde\theta}{2}\right), \quad 
    B = R_y\left(\frac{-\tilde\theta}{2}\right)R_z\left(\frac{-\tilde\alpha - \tilde\beta}{2}\right),\quad
    C = R_z\left(\frac{\tilde\beta -\tilde\alpha}{2}\right)
\end{equation}
The new angles are given by

\begin{align}    
\tilde \theta &= 2\arctan \left(\frac{|\beta^*\gamma - \alpha\eta^*}{|\beta\eta^* + \alpha^*\gamma|}\right),\notag \\
\tilde \alpha &= \arctan(\beta\eta^* + \alpha^*\gamma) + \arctan(\beta^*\gamma -\alpha\eta^*, \\
\tilde\beta &= \arctan(\beta\eta^* +\alpha^*\gamma) - \arctan(\beta^*\gamma - \alpha\eta^*) \notag
\end{align}

\end{widetext}
\end{document}